\documentclass[fleqn,usenatbib]{mnras}

\usepackage{newtxtext,newtxmath}
\usepackage{threeparttable}
\usepackage{academicons}
\DeclareRobustCommand{\VAN}[3]{#2}
\let\VANthebibliography\thebibliography
\def\thebibliography{\DeclareRobustCommand{\VAN}[3]{##3}\VANthebibliography}

\usepackage[T1]{fontenc}
\usepackage{hyperref}
\usepackage{wrapfig}
\usepackage{amsfonts}
\usepackage{amsmath}
\usepackage{tablefootnote}
\usepackage{bm}
\usepackage{graphicx}
\usepackage{epsfig}
\usepackage{mathtools}
\usepackage{color}
\usepackage{float}
\usepackage{lipsum}
\usepackage{booktabs}
\usepackage{orcidlink}
\usepackage{placeins}
\usepackage{graphicx}	% Including figure files
\usepackage{amsmath}	% Advanced maths commands
\usepackage[capitalize]{cleveref}
\usepackage{breqn}
\usepackage[mathscr]{euscript}
\usepackage{gensymb}
\usepackage{cuted}
\usepackage{widetext}
\usepackage{enumitem}
\usepackage{minibox}
\usepackage{multicol}

\newcommand{\pplus}{Pantheon$+$}
\newcommand{\LA}{\Lambda}
\newcommand{\LCDM}{$\LA$CDM}
\newcommand{\lcdm}{spatially flat $\LA$CDM}

\newcommand{\common}{\pplus/JLA common subsample}
\newcommand{\sne}{SNe~Ia}
\newcommand{\diff}{\mathrm{d}}
\newcommand{\prior}{\mathrm{Pr}}
\newcommand{\goesas}{\mathop{\sim}\limits}
\newcommand{\finity}{\mathop{\hbox{\it fi}}}
\newcommand{\mean}[1]{{\vphantom{\tilde#1}\bar#1}}
\newcommand{\OM}{\mean\Omega}
\newcommand{\Z}[1]{_{\lower2pt\hbox{$\scriptstyle#1$}}}
\newcommand{\X}[1]{_{\lower1pt\hbox{$\scriptscriptstyle#1$}}}
\newcommand{\Ns}[1]{_{\lower2pt\hbox{$\scriptstyle\rm#1$}}}
\newcommand{\ab}{\bar a} 
\newcommand{\rhb}{\bar\rho}
\newcommand{\bH}{\mean H}
\newcommand{\abn}{{\ab_0}}
\newcommand{\w}[1]{\mathop{\hbox{\,#1}}}
\newcommand{\lsim}{\lesssim}
\newcommand{\gsim}{\gtrsim}

\newcommand{\kms}{\w{km}\;\w{sec}^{-1}}
\newcommand{\kmsMpc}{\kms\w{Mpc}^{-1}}
\newcommand{\h}{\,h^{-1}}
\newcommand{\hM}{\h\w{Mpc}}

\newcommand{\ns}[1]{_{\rm #1}}
\newcommand{\OmLn}{\Omega\Z{\LA 0}}

\newcommand{\CC}{{c}}

\newcommand{\dd}{{\rm d}}

\newcommand{\Hn}{H\Z0}
\newcommand{\dL}{d\Z{L}}
\newcommand{\dA}{d\Z{A}}

\newcommand{\fv}{{f\ns v}}
\newcommand{\fvn}{f\ns{v0}}

\newcommand{\Hb}{\bH\Z{\!0}}
\newcommand{\zmin}{z\Ns{min}}
\newcommand{\zlo}{z\lsim\zmin}
\newcommand{\zhi}{z\gsim\zmin}
\newcommand{\HL}{Hubble--Lema\^{\i}tre}
\newcommand{\gb}{\mean\gamma}
\newcommand{\gw}{{\gb\ns w}}
\newcommand{\gv}{{\gb\ns v}}
\newcommand{\frn}[2]{\textstyle{\frac{#1}{#2}}}
\newcommand{\tb}{t'}

\newcommand{\mulcdm}{\mu\Z{\Lambda{\rm CDM}}}
\newcommand{\FF}{{\cal F}}
\newcommand{\zh}{\hat z}

\newcommand{\beq}{\begin{equation}}
\newcommand{\eeq}{\end{equation}}
\newcommand{\hblank}[1]{\hbox to#1 mm{\hfil}}
\newcommand{\hodge}{{*}}

{
    \definecolor{BLACK}{gray}{0}
    \definecolor{WHITE}{gray}{1}
    \definecolor{RED}{rgb}{1,0,0}
    \definecolor{GREEN}{rgb}{0,1,0}
    \definecolor{dgreen}{rgb}{.1,.6,.1}
    \definecolor{BLUE}{rgb}{0,0,1}
    \definecolor{CYAN}{cmyk}{1,0,0,0}
    \definecolor{MAGENTA}{cmyk}{0,1,0,0}
    \definecolor{YELLOW}{cmyk}{0,0,1,0}
    \definecolor{aw}{rgb}{0.2,0.5,0.75}
}
\hypersetup{
    colorlinks,%
    citecolor=blue,%
    filecolor=blue,%
    linkcolor=blue,%magenta
    urlcolor=blue
}

\title[Cosmology foundations revisited in Pantheon+]{Cosmological foundations revisited with Pantheon+}

\author[Lane et al.]{\fontsize{13pt}{16pt}\selectfont Zachary G.~Lane$^{1}$\orcidlink{0009-0003-8380-4003},
Antonia Seifert$^{1,2}$\orcidlink{0009-0005-9892-3667}, 
Ryan Ridden-Harper$^{1}$\orcidlink{0000-0003-1724-2885}, and
David L.~Wiltshire$^{1}$\orcidlink{0000-0003-1992-6682}\thanks{Corresponding author\newline ZGL and AS contributed equally to this work.}
\\
$^{1}$School of Physical and Chemical Sciences — Te Kura Mat\={u}, University of Canterbury, Private Bag 4800, Christchurch 8140, New Zealand\\
$^{2}$Institut f\"ur Theoretische Physik, Universit\"at Heidelberg, Philosophenweg 12, D-69120 Heidelberg, Germany\\
}

\date{Accepted 2024 October 23. Received 2024 October 7; in original form 2023 November 2}

\pubyear{2024}

\begin{document}
\label{firstpage}
\pagerange{\pageref{firstpage}--\pageref{lastpage}}
\maketitle

\begin{abstract}
We reanalyse the Pantheon+ supernova catalogue to compare a cosmology with non-FLRW evolution, the timescape cosmology, with the standard $\Lambda$CDM cosmology. To this end, we analyse the Pantheon+ for a geometric comparison between the two models. We construct a covariance matrix to be as independent of cosmology as possible, including independence from the FLRW geometry and peculiar velocity with respect to FLRW average evolution. This framework goes far beyond most other definitions of model independence. We introduce new statistics to refine Type Ia supernova (\sne) light-curve analysis. In addition to conventional galaxy correlation functions used to define the scale of statistical homogeneity we introduce empirical statistics which enables refined analysis of the distribution biases of \sne\ light-curve parameters ${\beta}c$ and $\alpha{x\Z1}$. For lower redshifts, the Bayesian analysis highlights important features attributable to the increased number of low-redshift supernovae, the artefacts of model-dependent light-curve fitting and the cosmic structure through which we observe supernovae. This indicates the need for cosmology-independent data reduction to conduct a stronger investigation of the emergence of statistical homogeneity and to compare alternative cosmologies in light of recent challenges to the standard model. Dark energy is generally invoked as a place-holder for new physics. {For the first time, we find evidence that the timescape cosmology may provide a better overall fit than $\Lambda$CDM and that its phenomenology may help disentangle other astrophysical puzzles.}
Our from-first-principles reanalysis of \pplus\ supports future deeper studies between the interplay of matter and nonlinear spacetime geometry in a data-driven setting. \\
\end{abstract}

%%%%%%%%%%%%%%%%% BODY OF PAPER %%%%%%%%%%%%%%%%%%

\section{Introduction}\label{sec:intro}

Modern cosmology is entering a new era characterised by orders of magnitude increase in the volume, variety and precision of available data, from both ground--based and space--based telescopes, from one--off missions to new dedicated observatories operating over decades. We are now learning to grapple not only with the entire electromagnetic spectrum, but also with multi-messenger astronomy.

Our new windows on the Universe -- high--energy neutrinos and gravitational waves from binary mergers of neutron stars and black holes -- offer glimpses of potential new physics which cannot be easily guessed at from consideration of only electromagnetism, nuclear/particle physics or gravity in isolation. Vital clues to unravelling new mysteries occur not only when all three multimessenger channels are detected, but also when some channels are detected in coincidence but others are not. The absence of directly detected neutrinos from GW170817 \citep{GW_2017} -- to date the most closely studied single astronomical event ever -- is a case in point.

In contemplating new physics we face the challenge that unlike electromagnetism which is linear,
non--Abelian gauge theories (nuclear physics) and general relativity (gravity) are each fundamentally nonlinear. This paper presents a data--driven detailed statistical reanalysis of the best available Type Ia supernovae (\sne) data \citep{Scolnic_2022}. However, rather than seeking new physics amongst the nonlinearities of either particle gauge theories or of general relativity (GR) alone, we will compare standard cosmology with a model that combines the nonlinearities of both key theoretical approaches.

This latter model, the timescape cosmology \citep{Wiltshire_2007_clocks,Wiltshire_2009_obs}, considers the fundamental questions, such as ``what is the origin of scale?'', ``where is infinity?'', and ``how do we compare standard to standardisable rulers and clocks?''. {It addresses the phenomenology commonly attributed to the dark sector or to modified gravity, but is} based on theoretical assumptions which are far more conservative than those commonly proposed to resolve cosmological tensions \citep{Di_Valentino_2021}. {Rather than modifying the Einstein-Hilbert action, the timescape model modifies the geometry of the background universe consistently with general relativity.}

\sne\ were the crucial dataset that first convinced \citep{Riess_1998, Perlmutter_1999} a majority of the community of the (apparently) accelerating rate of present epoch cosmic expansion. After more than two decades \sne\ are still one of the most widely investigated and crucial astronomical sources for determining geometric distances in the Universe. At a time when significant improvements in precision are now turning tensions into anomalies \citep{Peebles_2022}, we present a re-examination of \sne.

In the present analysis, we revisit the relationship of the foundations of geometry to the calibration of measurements in observational cosmology. In particular, 
a common usage of the phrase {``model independence''} is broadened to include independence from geometrical assumptions. Here we present our technical results alongside the underlying theories, to address numerous research communities engaged in studying the evolution of the Universe. We will examine how model-dependent assumptions potentially impact inherent \hbox{biases} in the standardisation of astrophysical data and model selection. This paper thereby introduces critical philosophical and scientific \hbox{questions} that the scientific community must address.

\subsection{$\Lambda$CDM, Timescape and Inhomogeneous Cosmology}\label{sec:cosmo}

The Friedmann-Lema\^{\i}tre-Robertson-Walker (FLRW) metric is built on two main assumptions, spatial isotropy and homogeneity, and it has been at the forefront of modern cosmology since the 1930s. Isotropy and homogeneity are supported by measurements of the Cosmic Microwave Background (CMB). The CMB is observed to have a near-perfect blackbody spectrum up to fluctuations of order $\vert \delta T \vert/T \goesas 10^{-3}$ \citep{Corey_1976, smoot_1977, cheng_1979}. The leading dipole fluctuation, with $|\delta T|=(3362.08 \pm 0.99)\,$~$\mu$K \citep{planck_2018}, is usually interpreted \textit{kinematically} as being purely due to our local motion with respect to a rest-frame.

With the leading dipole subtracted, the remaining anisotropies, with amplitude $\vert \delta T \vert/T \goesas 10^{-5}$ \citep{planck_2018}, are due to intrinsic density fluctuations on the surface of last scattering, or to inhomogeneities from structures on our line of sight. The small amplitude of the anisotropies is taken as a support for the \textit{Cosmological Principle}. While this is well justified on the surface of last scattering itself, the late epoch Universe contains a vast cosmic web of complex structures, requiring the fundamental understanding of average homogeneity and isotropy to be more nuanced \citep{Aluri_2022}. 

The standard cosmology is based on a FLRW {\em background} metric whose spatial curvature is found to be negligible \citep{planck_2018,Pierpaoli_2000}, i.e., the line element
\begin{equation}\label{eq:flrw}
    \diff s^2\Ns{FLRW} = - \diff t^2 + a^2(t) \left(\diff r^2 + r^2 \ \diff \Omega^2\right)
\end{equation}
is adapted to comoving observers at fixed $(r,\theta,\phi)$ who measure cosmic time, $t$, as their proper time. Here $\diff \Omega^2\equiv\dd \theta^2+\sin^2\theta\,\dd\phi^2$ is the metric on the celestial 2-sphere.
The word {\em background} contains an implicit assumption of a top--down ontology \citep{Wiltshire_2011}: the average evolution is assumed to contain the particular split of space and time implicit in the FLRW geometry regardless of the scale to which such an average is applied.  

For averages over very large spatial scales, linear perturbation theory on a global FLRW background is assumed to still apply today, at late cosmic epochs. When gravitational instability results in small-scale spatial structures forming, the inhomogeneities of standard cosmology enter the {\em nonlinear regime}, which must be treated by numerical simulations. Although simulations using full GR have been developed recently \citep{Giblin_2015, Bentivegna_2015, macpherson_2017}, most large numerical simulations apply Newtonian gravity to a simulation volume whose expansion is fit to a background spatially flat FLRW model.

In the standard approach, which envisages Hubble expansion as a global Hubble flow, all deviations from its average can always and everywhere be described by local Lorentz boosts, or \textit{peculiar velocities}. Such a situation is not typical of generic models in general relativity, including inhomogeneous dust cosmologies, which may exhibit non-kinematic differential expansion \citep{Bolejko_2016}.

With the standard kinematic interpretation of the CMB dipole, deviations from isotropy on the largest distance scales -- e.g., in galaxy and quasar distributions -- are best characterised in terms of the special relativistic aberration and modulation effects due to our local boost relative to the CMB rest-frame \citep{Ellis_1984}.

The kinematic interpretation is now challenged, however, at more than five standard deviations \citep{Secrest_2022, Secrest_2021}. Building on an earlier result,\footnote{\citet{Secrest_2021} use a sample of 1.36~million very distant quasars (with a mean redshift of 1.2) alone to find 4.9$\sigma$ tension. While the size of the quasar catalogue has increased, statistically the most significant aspect of the \citet{Secrest_2022} result are the {\em correlated} features in the smaller {\em independent} radio galaxy catalogue. Since the \citet{Secrest_2022} catalogue extends to fainter fluxes, the median redshift should be slightly greater than the mean stated by \citet{Secrest_2021}, and is not relevant for this analysis. The radio galaxies of \citet{Secrest_2022} have a median redshift of 0.8 \citep{Condon_1998}.} \citet{Secrest_2022} combine a larger quasar catalogue of 1.6~million source with a sample of 508,144 not--quite--so distant radio galaxies to find a 5.1$\sigma$ disagreement with the kinematic interpretation. This anomaly now exceeds the $5\sigma$ threshold used to demarcate signals worthy of {new physics} attention in high--energy physics accelerator experiments -- as has been noted by careful readers \citep{Peebles_2022}. 
%% Philosophers - see also (i) https://ui.adsabs.harvard.edu/abs/2022APS..APRD02001P/abstract ; (ii) https://ui.adsabs.harvard.edu/abs/2022wtcr.book.....P/abstract

One must add the caveat that evolutionary effects in the quasar luminosity function lead to a variability of approximately $3\sigma$ in estimates of the amplitude for the Ellis-Baldwin dipole using current best astrophysical models \citep{Guandalin_2023}. However, by any measure it remains a serious tension. To address such challenges, in this paper we will broaden the definition of model--independence to include global non--FLRW models which agree with the standard cosmology to the extent that: 
\begin{enumerate}[label=(\roman*), topsep=0pt, itemsep=0pt,leftmargin=*, align=right]
    \item\label{num:highz} a notion of an isotropic average \HL\ expansion law exists on scales $\zhi$, where the redshift $\zmin$ is determined by a revised notion of {\em statistical homogeneity scale} (SHS);
    \item\label{num:lowz} on scales $\zlo$ the directly observed small--scale inhomogeneous matter distribution is produced via gravitational instability from primordial density fluctuations. Viability requires consistency with all observed phenomenology of the CMB radiation on one hand, and with nuclear physics models on the other. The latter have been well--established in terrestrial accelerator experiments at energies below that of the quark--gluon transition \citep{Rafelski_2020}.
\end{enumerate}

Criterion \ref{num:highz} is addressed by general models that incorporate \textit{backreaction} of inhomogeneities \citep{Buchert_2000, Buchert_2001, Buchert_2020},\footnote{For general reviews see \citet{Buchert_2012, Wiltshire_2014_cosmic}.} namely additional terms in the averaged Einstein equations governing cosmic evolution on large scales. Such terms arise from the bottom up when coarse--graining Einstein's equations for structure on small scales.

Criterion \ref{num:lowz} is addressed by phenomenological models which calibrate radiation and matter transfer functions from the early Universe to non-FLRW evolution in the nonlinear regime\footnote{In contrast to electromagnetism, which is linear, non-Abelian gauge theories and general relativity are both nonlinear. As a gauge theory, electromagnetism involves a single complex scalar potential, $\psi$, and the internal symmetry group $U(1)$. Furthermore, in GR with the absence of torsion, covariant derivatives in Maxwell's equations reduce to ordinary exterior derivatives: $\dd {\mathbf F} = 0$, $\dd \hodge {\mathbf F} = \mu\Z0 {\mathbf J}$, where $\mathbf F=\dd {\mathbf A}$, $\mathbf A$ is {the 1-form corresponding to the 4-vector potential}, ${\mathbf A}=\dd\psi$, $\mathbf J$ {the current 1-form}, and $\hodge$ denotes the Hodge dual \citep{Szekeres_2004}.} at late epochs \citep{Nazer_2015, Racz_2017}. 

The timescape model uses a particular implementation of Buchert's averaging scheme with regional average metrics\footnote{Inspired by \citet{Ellis_1984_fitting}, in  timescape cosmology finite infinity regions are defined as (marginally expanding) boundaries of spatially flat regions containing bound structures. An alternative related quantity is the {\em matter horizon} \citep{Ellis_2009}.
The quantities $\tau_{\rm w,v}, \eta_{\rm w,v}, a_{\rm w,v}$ are not defined globally but can be adapted to the clocks and rulers of ideal regional observers. Quantities in finite infinity region, $\tau_{\rm w}, \eta_{\rm w}, a_{\rm w}$, differ from their counterparts in void domains, ${\cal D}\Z{\rm v}$, since finite infinity regions average over the many sources of gravitational energy within the largest bound structures (binding, rotational, thermal, \dots) whereas voids are dominated by gradients in the kinetic energy of expansion. {In the timescape model, apparent cosmic acceleration} results from a cumulative misidentification and miscalibration of gradients in the quasilocal kinetic energy of cosmic expansion as structures grow. See \citet{Wiltshire_2011,Wiltshire_2014_cosmic} for further introduction to these ideas, and \cite{Dam_2017, Buchert_2020} for later updates.
}
\begin{equation}\label{eq:finitymetric}
    \diff s^2_{\finity} = - \diff \tau_{\rm w}^2 + a_{\rm w}^2(\tau_{\rm w}) \left(\diff \eta_{\rm w}^2 + \eta_{\rm w}^2 \ \diff \Omega^2\right),
\end{equation}
for \textit{finite infinity} regions within the walls and filaments of the cosmic web, and
\begin{equation}\label{eq:voidmetric}
    \diff s^2_{{\cal D}\Ns{\rm v}} = - \diff \tau^2_{\rm v} + a_{\rm v}^2(\tau_{\rm v}) \left(\diff \eta_{\rm v}^2 + \sinh^2(\eta_{\rm v}) \ \diff \Omega^2\right),
\end{equation}
at void centres \citep{Wiltshire_2007_clocks, Wiltshire_2008, Wiltshire_2009_obs}.

Both of these metrics have FLRW form: \cref{eq:finitymetric} is identical to \cref{eq:flrw} and the void metric \cref{eq:voidmetric} is a negative spatial curvature FLRW metric. However, in the timescape model these metrics are {\em not} exact solutions of the Einstein equations in general. Rather, they are regional best-fit effective average metrics obtained in one particular solution \citep{Wiltshire_2007_clocks,Wiltshire_2007_sol} to the {\em fitting problem} \citep{Ellis_1984_fitting, Ellis_1987}. The effective metrics apply to small regions within which the Einstein equations hold to a good approximation, and which are in turn embedded in the global average geometry of the complex cosmic web.

Applying Buchert's scalar averaging formalism to the geometry, \cref{eq:finitymetric,eq:voidmetric}, of an ensemble of wall and voids yields the \textit{bare cosmological parameters} \citep{Wiltshire_2007_clocks, Wiltshire_2009_obs}
\begin{align}
    \OM_{\rm M} &\equiv \frac{8\pi G\rhb\Ns{M0}\abn^3}{3\bH^2\ab^3}\,, & 
    \OM_{\rm R} &\equiv \frac{8\pi G\rhb\Ns{R0}\abn^4}{3\bH^2\ab^4}\,,\label{eq:omr}\\
    \OM_{\rm K} &\equiv \frac{-k\Ns{v} f_{{\rm v}i}^{2/3} f\Ns{v}^{1/3}}{\ab^2\bH^2}\,, &
    \OM_{\cal Q} &\equiv \frac{-\dot f_{\rm v}^2}{9f\Ns{v}(1-f\Ns{v})\bH^2}\,,\label{eq:om}
\end{align}
in terms of a {\em volume-average comoving time}, $t$; the average expanding volume, ${\cal V}\propto{\ab}^3$ with $\ab^3 = f\Z{{\rm v}i} a_{\rm v}^3 + (1 - f\Z{{\rm v}i}) a_{\rm w}^3$ on effective spatial {hypersurfaces}, $t=\w{const}.$; the bare Hubble parameter, $\bar H=\dot\ab/\ab$; the \textit{void volume fraction} $f\Z{\rm v}(t) = f\Z{\mathrm{v}i} a_{\rm v}^3 / \ab^3$ with initial value $f\Z{{\rm v}i}$ at last scattering; $k_{\rm v}=\w{const}$;
and the subscript $0$ denotes present epoch values of the average energy densities of nonrelativistic matter, $\bar\rho \Ns M$; radiation, $\bar\rho \Ns R$; etc.

An average global Hubble-Lema\^{i}tre expansion law parameterised by $\bH(t)$ only applies on scales $\gsim 70$--$120\hM$ \citep{Hogg_2005, Scrimgeour_2012}, and the transition to homogeneity is conventionally defined via density contrasts, $\delta_\rho\equiv(\rho-\bar\rho)/\bar\rho$, relative to a global FLRW background \citep{Gabrielli_2005}. To circumvent this for non-FLRW models, the {\em Statistical Homogeneity Scale} (SHS) is introduced as a neologism\footnote{Inventing modern neologisms concerning the nature of space and time involves new and different challenges to even the ones Einstein faced. Further philosophical discussions, aside from the Conclusion (\cref{sec:conclusions}), are beyond the scope of this paper and will be published elsewhere. Since extraordinary claims require extraordinary evidence, when faced with a genuine paradigm shift, the responsibility of any scientist to actively seek to ``kill'' their own theory is even greater. The directive given by DLW to the coauthors of this article was to apply precisely such an approach.} for the scale determined by $n$-point galaxy correlation functions.

In Buchert averages, volume-average parameters defined as in \cref{eq:om,eq:omr} obey a sum-rule\footnote{An inhomogeneous $\OM\Z{\Lambda}$  parameter, induced by a global cosmological constant, can also be added mathematically \citep{Buchert_2020}. While of formal interest, it is not part of timescape phenomenology and is not considered here. In addition, \citet{Buchert_2020} also introduce terms for dynamical backreaction, $\OM\Ns{\cal P}$, and stress-energy backreaction, $\OM\Ns{\cal T}$, in the full generalisation of \cref{eq:sum_rule}. Such terms are likely to be important in the very early Universe, but not at late cosmic epochs when non-relativistic matter dominates, and when $|\OM\Ns{\cal P,T}|\ll|\OM\Ns{\cal Q}|<0.044$. The latter inequality is a strict upper bound in the timescape model but not in generic backreaction schemes. We also use $\OM\Ns{K}$ rather than the alternative $\OM\Ns{\cal R}$ to avoid confusion with the radiation density parameter, $\OM\Ns{R}$, which is significant in the early Universe.}
\begin{equation}\label{eq:sum_rule} 
\OM\Ns{\rm M} + \OM\Ns{\rm R} + \OM\Ns{\rm K} + \OM\Ns{\cal Q}=1, 
\end{equation}
similarly to the Friedmann equation. By contrast, $\OM\Ns{K}$ does not scale as a simple power of $\ab$. Instead, average spatial curvature is directly related to the void fraction: $\OM\Ns{K}\propto f_{\rm v}^{1/3}/(\ab^2\bar H^2)$. Here $f_{\rm v}\to 0$ at early times but dominates at late epochs. The {\em kinematical backreaction} term $\OM \Ns{\cal Q}$ gives rise to non-FLRW evolution if $\dot f_{\rm v}\ne0$.

Volume--average time, $t$, only coincides with proper time for ideal isotropic observers whose regional spatial curvature coincides with the global volume average.

Since the Universe is dominated in volume by voids, such locations are necessarily within voids, though not at their centres. Phenomenological lapse functions,
\begin{equation}\label{eq:pharlapse}
\gw\equiv\frac{\dd t}{\dd\tau_{\rm w}},\qquad \gv\equiv\frac{\dd t}{\dd\tau_{\rm v}}
\end{equation}
relate the proper times $\tau_{\rm w,v}$ to volume-average time, $t$. For observers and sources in bound structures that have broken away from the Hubble expansion only the wall lapse, $\gw$, is directly relevant. Structure formation thus gives rise to a selection effect: our mass-biased view of the Universe differs systematically from the volume-average.

Timescape phenomenology is based theoretically on an extension of the Strong Equivalence Principle to regional averages, via its {{\em Cosmological Equivalence Principle} (CEP) \footnote{For an introductory essay on the CEP see \citet{Wiltshire_2009_fqxi_essay}.} \citep{Wiltshire_2008} which encodes a} {\em uniform quasilocal Hubble expansion condition}. This is used to conformally match the regional geometry (\cref{eq:finitymetric}) to the non-FLRW geometry along the past light cone. One then arrives at effective dressed cosmological parameters, including, e.g., the dressed matter density with present epoch value
\begin{equation}\label{eq:omegam_ts}
    \Omega\Ns{{\rm M}0}\equiv\left.\gw^3 \bar\Omega \Ns{\rm M}\right|_0 = \frac12 (1 - f\Ns{v0}) (2 + f\Ns{v0}).
\end{equation}
Dressed parameters do {\em not} obey a sum rule akin to \cref{eq:sum_rule} for the bare parameters, and cannot be interpreted in the conventional FLRW manner.
Nonetheless, since $\Omega\Ns{{\rm M}0}$ takes similar numerical values to the FLRW matter density parameter, its use is convenient for plotting timescape parameter constraints alongside those for \lcdm. 

{Just as Einstein's statements of the equivalence principles preceded decades of inquiry and debate about their full mathematical implementation, in addressing the question ``what is the origin of scale?'', the CEP provides a foundation for a new observational phenomenology that might serve in the quest for a quantum theory of gravity. Our current understanding of the mechanism by which massless particles gain mass in the early Universe -- the Higgs mechanism -- is purely field theoretic, without reference to gravity. Our best attempts to understand quantum gravity, including string theory, invoke extra spatial dimensions to unify particle physics and geometry. Since processes in the very early Universe laid the seed fluctuations for the observed matter power spectrum, observational cosmology may provide a window to understanding the geometry of the Higgs mechanism, if it is to be found. As argued elsewhere \citep{Coley_2017} the quasilocal uniform Hubble expansion condition is consistent with a number of approaches to quantum gravity. At the very least, it provides a conceptual and phenomenological framework within which ideas different to the FLRW assumption may be tested.

By providing a mathematical closure condition to the Buchert equations, thereby determining how much initial fluctuations can grow from the very early Universe via purely causal processes, the CEP provides an explanation for the spatial isotropy of distant sources in large full sky surveys when averaged on our celestial sphere. This contrasts strongly with the standard cosmology in which average spatial isotropy (and homogeneity) are assumed.}

Among non-FLRW models with backreaction of inhomogeneities, the timescape model is observationally the most well-developed. Most significantly, in {17} years since its expansion history was first constrained by \sne\ data \citep{Leith_2008}, the timescape still stands as a viable alternative to \LCDM, offering a foundational explanation for {the expansion history commonly attributed to dark energy}. In addition to being well-supported in early studies of gamma-ray-burst distance indicators \citep{Smale_2011_grb} and ongoing \sne\ analyses \citep{Dam_2017}, constraints have also been derived from the CMB \citep{Nazer_2015} using Planck data \citep{Planck_2013}, and the Baryon Acoustic Oscillation signal has been extracted directly \citep{Heinesen_2019} from Baryon Oscillation Spectroscopic Survey (BOSS) data \citep{Alam_2017}. {Most recently, a reanalysis of dimensionless lens/source distance ratios \citep{HarveyHawes_2024} determined from the largest available strong gravitational lensing catalogue \citep{Chen_2019} independently picks out the same value of the free parameter, the void volume fraction.}

{Void statistics have now been studied for the first time in large cosmological simulations using full general relativity constrained only by the initial matter power spectrum \citep{Williams_2024}. The results are quantitatively consistent with the timescape model. Specifically, by adapting standard watershed void finders to the {\sc Einstein Toolkit}, \citet{Williams_2024} have identified the signature of emerging spatial curvature common to all phenomenologically realistic backreaction scenarios \citep{Wiltshire_2009_obs,Roukema_2013,Bolejko_2018}. Further, the magnitude of the kinetic spatial curvature is consistent with timescape parameters fit to Planck CMB data \citep{Duley_2013}. Detailed analysis of voids in the numerical relativity simulations provides direct insights into how the monopole celestial sky average of the kinetic curvature is interpreted as dark energy in the standard cosmology \citep{Wiltshire_2024}. This further explains the link to the weak field $N$-body code {\it gevolution} \citep{Adamek_2016}, which operates by defining a renormalised scale factor and a renormalized time coordinate when subtracting a monopole to retain an unperturbed Friedmann metric as the spatial average. }

Since \lcdm\ has been a good phenomenological fit to independent datasets for two decades, any viable contender to displace the standard cosmology must necessarily produce many results which are very close. An early study \citep{Wiltshire_2009_obs} thus concentrated on quantifying the precision needed to distinguish timescape from \LCDM: via the Clarkson-Bassett-Lu (CBL) test \citep{Clarkson_2008}, the $Om(z)$ test \citep{Sahni_2008,Shafieloo_2009}, the Alcock--Paczy\'nski test \citep{Alcock_1979} and the redshift-time drift test \citep{Sandage_1962}. Among these observational tests, the CBL test is perhaps the most significant as it yields a test statistic $\Omega\Ns{k}(z)$ of the Friedmann equation: $\Omega\Ns{k}(z)$ is constant with redshift, and highly constrained by the CMB, for any FLRW expansion history whether embedded in an exotic matter model or in a modified gravity model. Any non-FLRW expansion history will differ from this. Timescape yields differences of $1$--$3$\% in the distance--redshift relation at any finite $z$, leading to large cumulative correlated differences over a long lever arm in the expansion history. Timescape was used for independent projections for the Euclid mission in 2014 \citep{Sapone_2014}. Consequently, now that Euclid has finally been launched, in the next six years timescape itself will be tested as will any FLRW expansion history, including theories routinely studied by the majority of the community \citep{Di_Valentino_2021}. 

The projections made a decade ago \citep{Sapone_2014} yield a potential falsification of the Friedmann equation based on robust BAO extraction by Euclid, and independent distances and redshifts for 1000 \sne. The biggest systematic uncertainties that the small inhomogeneous cosmology community has always found the most daunting relate to the use of the FLRW expansion history in data reduction pipelines\footnote{Adapting matter and radiation transfer functions from the early Universe to the present epoch requires a proxy for the small but nontrivial violation of the Friedmann equation at last scattering. Timescape provides a quantitative framework for this, with initial values $f_{{\rm v}i}\goesas2\times10^{-5}$, $\bar\Omega_{Ki}\goesas7\times10^{-5}$ and $\bar\Omega_{{\cal Q}i}\goesas-1\times10^{-5}$ at decoupling in \cref{eq:sum_rule}. Nonetheless, by adapting {\sc class} \citep{Lesgourgues_2011, Blas_2011}
to match perturbed FLRW models to timescape \citet{Nazer_2015} found that the range of physically viable replacements for $\bar\Omega_{{\cal Q}i}$ led to $8$--$13$\% systematic uncertainties in present epoch timescape model parameters \citep[Table 1]{Duley_2013}.}. With the most recent \pplus\ data release \citep{Scolnic_2022}, we are now in a position to directly test the \sne\ expansion history projection made a decade ago by \citet{Sapone_2014} by bootstrap resampling. In doing so, we will also directly perform Bayesian tests that not only compare the \lcdm\ model with the timescape model, but which have a very wide impact in broadening criteria for model independence in \sne\ analysis.

\subsection{Supernova Redshift-Distance Analysis}\label{sec:sne}

\citet{Dam_2017} investigated the redshift -- luminosity distance data of Type Ia supernovae (\sne) using the \textit{Joint Light-curve Analysis} (JLA) \sne\ catalogue \citep{Betoule_2014} to determine the likelihood of \lcdm\ and timescape. This analysis found no statistical preference favouring either model. The JLA sample consists of 740 unique supernovae and was the largest \sne\ catalogue available at the time of the \citet{Dam_2017} analysis. The JLA catalogue has since been superseded by \textit{Pantheon} \citep{Scolnic_2018} and most recently \textit{\pplus} \citep{Scolnic_2022}. The full \pplus\ sample contains a dataset of 1708 supernova from 20 different surveys and  1550 unique supernovae. The \pplus\ and JLA samples have 584 supernovae in common.

The remaining 176 supernovae of the JLA catalogue have not been included in \pplus\ as \citet{Scolnic_2022} imposed quality cuts on this survey, partly to satisfy stricter requirements on pre-maximum luminosity data. Specifically, \citet{Brout_2022_cosmo} identify significant differences in the distance modulus for SNLS and DES supernovae with redshifts exceeding 0.8. These disparities are contingent upon the inclusion of the $U$~band data at lower redshifts, which is incorporated in their SALT2 training set \citep{Scolnic_2022}. The \sne\ relying on this data are removed from the dataset due to the incompatibility for cross-calibration.

Both the JLA and \pplus\ use the SALT2 algorithm \citep{Guy_2007} to standardise \sne\ light-curves for cosmological analysis. For each light-curve SALT2 fits the time stretch parameter ($x\Z 1$), colour ($c$), and amplitude $x\Z 0$, which is then used to calculate the rest-frame apparent magnitude of the $B$~band filter ($m\Ns{B}=-2.5\log_{10}(x\Z 0)$). These parameters depend on the SALT2 surfaces used, which were updated between JLA and \pplus. The differences arising from the update are discussed in \citet{Taylor_2021} and result in differences between the common subsample of \sne.

The SALT2 parameters can be combined into the distance modulus $\mu$ via the Tripp formula \citep{Tripp_1998},
\begin{equation}\label{eq:tripp}
    \mu = m - M = m\Ns{B}^* - M\Ns{B} + \alpha x\Z 1 - \beta c,
\end{equation}
where $m\Ns{B}^*$ denotes the apparent magnitude at maximum in the rest-frame $B$~band and $M\Ns{B}$ is the corresponding absolute magnitude of the source. $\alpha$ and $\beta$ are assumed to be constant, \citet{Scolnic_2022} adopt values of $0.148$ and $3.112$, respectively, for their nominal fit. Since \citet{Scolnic_2022} calculate their nominal fit with $w$CDM and we wish to conduct a model-independent analysis similar to that of JLA, we do not use these values.

\begin{figure}
    \includegraphics[width=\columnwidth]{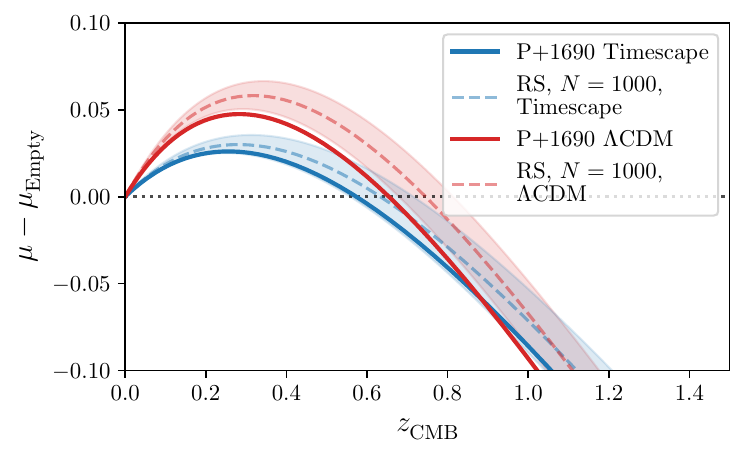}
    \caption{The residual distance moduli for the P$+1690$ sample following $\mu\Ns{\Lambda CDM}(z\Ns{CMB}) - \mu\Ns{Empty} (z\Ns{CMB})$ and $\mu\Ns{TS}(z\Ns{CMB}) - \mu\Ns{Empty}(z\Ns{CMB})$, calculated with $\Hn = 66.7\,\rm kms^{-1}\,Mpc^{-1}$ and assuming the best-fitting parameters at $z\Ns{CMB} = 0.033$: $\Omega\Ns{M0} = 0.433$ for \lcdm\ and $f\Ns{v0} = 0.6751$ for timescape~(TS). For this dataset, the maximum difference of the distance moduli in the $\mu\Ns{TS}(z\Ns{CMB}) < \mu\Ns{\Lambda CDM}(z\Ns{CMB})$ region, i.e. $z\Ns{CMB} < 0.8850$, is $\mu\Ns{\Lambda CDM}(z\Ns{CMB}) - \mu\Ns{TS}(z\Ns{CMB}) = 0.0222$ at $z\Ns{CMB} = 0.3375$. The median and the standard deviation for random subsamples (referred to as RS) of size $N = 1000$ (obtained as described in \cref{sec:random}) are shown for reference.}
    \label{fig:mu_empty}
\end{figure}

The observational distance modulus serves as a model for band correction
\begin{equation}
    \Delta \mu\Ns{B} \equiv (m - M) - (m\Ns{B}^* - M\Ns{B}) = \alpha x\Z 1 - \beta c, \label{eq:bandaid}
\end{equation}
to the theoretical distance modulus, shown by
\begin{equation}
    \mu \equiv 25 + 5 \log_{10}\left(\frac{\dL}{\rm Mpc}\right)\label{eq:mu},
\end{equation}
which is determined using the bolometric flux. The luminosity distance $d_L$ can be calculated using the redshift and suitable model parameters $\Omega\Ns{M0}$ and $\Omega\Ns {\Lambda0} = 1 - \Omega\Ns{M0}$ for the \lcdm\ model and $f\Ns{v0}$ for the timescape cosmology. For details on this calculation, see \cref{app:model} and \citet[Appendix~A]{Dam_2017}. Evaluating the distance moduli relative to an empty \textit{Milne} universe (\cref{fig:mu_empty}) reveals the key differences between the models tested here. Since the observed cosmic acceleration is an apparent phenomenon in the timescape model, the distance modulus values, denoted as $\mu\Ns{TS}$, fall between those of the $\mu\Ns{\Lambda CDM}$ model and the empty case. However, it is worth noting that they are positioned closer to the $\mu\Ns{\Lambda CDM}$, as illustrated in \cref{fig:mu_empty} and \citet[Fig.~1]{Dam_2017}.

Another approach used extensively in the 1990s--2000s to `standardise' \sne\ light-curves is the \textit{Multi-colour Light Curve Shape} (MLCS) method, along with other empirical methods. These methods were pivotal in the breakthrough discovery by \citet{Riess_1998} and \citet{Perlmutter_1999}. MLCS is an elaborate empirical model designed to correct the absolute magnitude and colour of \sne\ light-curves, rendering them suitable as {\textit{standardisable candles}}. 

One notable distinction between SALT and MLCS2k2 (an updated version of MLCS) lies in their implementation of $K$-corrections \citep{Jha_2007}. In the case of SALT, these corrections are integrated into the model, while MLCS2k2 determines the $K$-corrections more empirically through coefficient modelling, with uncertainties derived from measurement errors \citep{Guy_2005}. The approach in MLCS2k2 involves modelling multi-colour light-curves separately to calculate luminosity distances, as opposed to estimating model parameters.

Several authors have observed that SALT2 and MLCS2k2 yield different results in the computed distance modulus. The differing results are associated with how each method corrects for and handles systematic uncertainties \citep{Kessler_2009_Sloan, Hicken_2009}. SALT2 and MLCS2k2 introduce the dependence on the cosmological model at different stages of the analysis and data reduction. In MLCS2k2, the cosmological parameters are not entangled with the distance modulus, whereas SALT2 utilises cosmological parameters to estimate a distance modulus, thereby maintaining a degree of model dependence \citep{Kessler_2009_Sloan, Hicken_2009, Smale_2011_sne}.

As demonstrated in the study by \citet{Smale_2011_sne}, differences between various datasets reduced using MLCS2k2 and SALT2 are significant, especially in the context of the timescape model. Even when SALT2 is applied to the timescape model using \texttt{simple_cosfitter}, the light-curve parameters provided have still been fitted with the standard cosmology. However, with the rigorous approach that \pplus\ takes with SALT2 we are able to extract the relevant parameters before cosmological biases are introduced, which is crucial for this analysis. 

In this paper, we use the \pplus\ supernova catalogue to conduct a fair and more detailed comparison of both the \lcdm\ and timescape models. To this end, we construct suitable input and covariance files as described in \cref{sec:input}. The methods used for the statistical analysis are explained in \cref{sec:statmethods}. In \cref{sec:results}, we analyse the light-curve parameters found from the \pplus\ sample by our statistical approach and compare the different cosmological models. We continue with a discussion of the most relevant findings in \cref{sec:discussion} and present our conclusions in \cref{sec:conclusions}.

\section{Pantheon+ Data}\label{sec:input}

The \pplus\ data release \citep{Scolnic_2022} includes the heliocentric and CMB corrected redshifts, the light-curve parameters $m\Ns{B}$, $x\Z 1$, and $c$ alongside the covariances of the light-curve parameters for all observed supernovae. The CMB corrected redshifts given by \pplus\ {\citep{Peterson_2022}} include a correction for the peculiar velocity field, which implicitly assumes the standard cosmological model based on the FLRW metric and Newtonian $N$-body simulations. As we test a fundamentally different cosmology, we apply a boost from the heliocentric rest-frame to calculate the redshift in the CMB rest-frame without such corrections. In particular, we adopt the \citet{Fixsen_1996} values: $(\ell, b) = (264.14\degree, 48.26\degree)$, $v\Ns{CMB} = 371.0\,$km\,s$^{-1}$, $\rm RA\Ns{CMB} = 168.01\degree$, and $\rm DEC\Ns{CMB} = -6.983\degree$. Consequently, we do not consider any covariance terms associated with the peculiar velocity treatment by \pplus\ either.

\citet{Brout_2022_cosmo} provide the results and the covariance matrix in terms of the distance modulus, which is determined by the SALT2mu algorithm from the \texttt{SNANA} package \citep{Kessler_2009}. This algorithm requires the input of a reference and fiducial cosmological model to the light-curve SALT2 fitting parameters\footnote{The results and the covariance are given in terms of $x_0$, $x\Z 1$ and $c$, and their covariance can be transformed as ${\rm Cov}(m\Ns{B}, x\Z 1, c) = \mathbb{J} \: {\rm Cov}(x\Z 0, x\Z 1, c) \: \mathbb{J}^\intercal$.} to calculate the distance modulus. As the reference cosmology used was $w$CDM, any analysis performed with these biases would inherently favour models based on the FLRW framework. Therefore, we do not account for several bias-corrections from the Hubble diagram analysis such as Malmquist bias or peculiar velocity for a fair model comparison. The reconstruction of the covariance matrix considering these changes with respect to the one given in \citet[Section~2.2]{Brout_2022_cosmo} is presented here in \cref{sec:statcov,sec:systcov}.
In doing so, we also consider the three light-curve parameters taken directly from SALT2 as separate input parameters, instead of combining them into a single distance modulus parameter, $\mu$. We extract input and covariance files in terms of $m\Ns{B}$, $x\Z 1$ and $c$, to replicate the methods performed by \citet{Dam_2017} and \citet{Nielsen_2016} with the JLA catalogue. In this paper, our reanalysis of JLA differs from the approach taken in \citet{Dam_2017} as we calculate the $\sigma^2\Ns{lens}$ and $\sigma^2_z$ {excluding the peculiar velocity correction terms}.

\subsection{Statistical Covariance}\label{sec:statcov}

Considering the input from the fitted light-curve parameters yields a $3 \times 3$ covariance block for each supernova. Combining all the $\mathcal{N} = 1708$ supernovae gives a $3\mathcal{N} \times 3\mathcal{N}$ block-diagonal matrix, which accounts for the SALT2 measurement uncertainty term $\sigma\Ns{meas}^2$ in the covariance matrix by \citet{Brout_2022_cosmo}. The redshift covariance $\sigma^2_z$ and the lensing term $\sigma^2\Ns{lens} = 0.055\,z\Ns{CMB}$ are calculated from the boosted CMB redshift and added to the $(m\Ns{B,i}, m\Ns{B,i})$ matrix elements. The $\sigma^2\Ns{floor}$ from the SALT2mu algorithm and $\sigma^2\Ns{vpec}$ terms are not accounted for, as they introduce a bias towards the standard cosmology based on the FLRW metric. The full statistical covariance is thus a $3 \mathcal{N} \times 3 \mathcal{N}$ block-diagonal matrix build from $\mathcal{N}$ blocks of the form
\begin{equation}\label{eq:statcov}
    \left(\begin{array}{ccc}
        (\Delta m\Ns{B})^2 & \Sigma_{m\Ns{B}, x\Z 1} & \Sigma_{m\Ns{B}, c} \\
        \Sigma_{m\Ns{B}, x\Z 1} & (\Delta x\Z 1)^2 & \Sigma_{x\Z 1, c} \\
        \Sigma_{m\Ns{B}, c} & \Sigma_{x\Z 1, c} & (\Delta c)^2
    \end{array}\right) + 
    \left(\begin{array}{ccc}
        \sigma^2_z + \sigma^2\Ns{lens} & & \\
        & 0 & \\
        &  & 0
    \end{array}\right),
\end{equation}
one for each supernova.

However, some of the underlying blocks from the light-curve fitting (left term of \cref{eq:statcov}) are not positive semi-definite. For the \pplus\ analysis this is not a problem, since their full statistical covariance matrix includes more terms which create a positive semi-definite matrix. As our formulation of the covariance matrix differs, the complete covariance matrix calculated is not positive semi-definite. To rectify this, we drop 15 supernovae that have non-positive semi-definite statistical $3 \times 3$ covariance matrices (see \cref{tab:dropped}) to get only non-negative real eigenvalues for the full statistical covariance. The resulting nominal sample consists of $\mathcal{N} = 1690$ supernovae, $N = 1535$ of which are unique, and will be called \textit{P+1690}. For use in future catalogues (\cref{sec:modelcomp} below) the maximum number of unique \sne\ events will be denoted $N_{\rm u}$; here $N_{\rm u} = 1535$.

\subsection{Systematic Covariance}\label{sec:systcov}

The systematic covariance by \citet{Brout_2022_cosmo},
\begin{equation}\label{eq:systcov}
    C\Ns{syst}^{ij} = \sum_\psi \sigma \Z \psi \partial \Z \psi \mu^i \partial \Z \psi \mu^j,
\end{equation}
accounts for the different calibration uncertainties of the light-curve parameters and models by iterating over all different systematics ($\psi$) mentioned in \citet[Table~4]{Brout_2022_cosmo} and assigning respective scale factors $\sigma 
\Z \psi$. Here, $\partial \Z \psi \mu = \mu \Z \psi - \mu\Ns{nominal}$ is the deviation caused by each calibration. This gives a $\mathcal{N} \times \mathcal{N}$ matrix for the systematic covariance associated with the distance modulus $\mu$. 

As the same uncertainties apply to our dataset, we calculate our systematic covariance similarly; but do not account for any calibration associated with peculiar velocities. Furthermore, we do not account for deviations in $\mu$ only, but consider the three parameters $m\Ns{B}$, $x\Z 1$ and $c$ separately\footnote{\citet{Brout_2022_cosmo} consider $\partial \Z \psi \Delta \mu = \Delta\mu \Z \psi - \Delta\mu\Ns{nominal}$ with $\Delta \mu = \mu - \mu\Ns{ref}$ in \cref{eq:systcov}, where $\mu\Ns{ref}$ is the result of their SALT2mu algorithm and thus assumes a standard FLRW-based cosmology framework implicitly. However, as $\mu\Ns{ref}$ does not depend on the calibration systematics $\psi$ except for peculiar velocity calibrations (which we neglect), it drops out of the equation in our case. We do not consider reference values for any of the three light-curve parameters for this reason. {Peculiar velocity corrections can be introduced for the comparison of \LCDM\ with other FLRW models, however, a $\mu\Ns{ref}$ will have to be provided.}}. Thus, instead of running over the supernova events only, $\partial \Z \psi \mu^i \in \{\partial \Z \psi \mu^1, ..., \partial \Z \psi \mu^{\cal N}\}$, we consider $\partial \Z \psi \tilde\mu^i \in \{\partial \Z \psi m\Ns{B}^1, \partial \Z \psi x\Z 1^1, \partial \Z \psi c^1, ..., \partial \Z \psi m\Ns{B}^{\cal N}, \partial \Z \psi x\Z 1^{\cal N}, \partial \Z \psi c^{\cal N}\}$ and obtain a $3 \mathcal{N} \times 3 \mathcal{N}$ matrix in this way.

By constructing the covariance matrix following the methodology employed by \citet{Brout_2022_cosmo}, we incorporate all of the known systematics in a manner consistent with the updated SALT2 algorithm. In particular, this includes the mass step \citep{Popovic_2021_sim} and redshift-dependent Hubble scatter \citep{Brout_2021}. Additional potential systematics, as outlined in \citet{Popvic_2021_ref} have to be investigated further before being incorporated into the algorithm. Presently, they are not included in the SALT2 treatment and as a result we do not take them into consideration at this stage.

\subsection{Duplicate Supernovae}

Observations of the same supernovae event from different surveys were included in \pplus\ as differing \sne. To account for covariances between surveys for a given supernova event, we determine a \textit{duplication} covariance,
\begin{equation}
    C_{{\rm dupl}, i}^{rs} = \delta \tilde\mu^r \delta \tilde\mu^s,
\end{equation}
for all duplicated \sne. The indices $r, s$ run over all surveys $1, ..., n \Z i$ that observed this particular supernova event. This matrix is constructed similar to \cref{eq:systcov}, but considers the deviation from the mean over all $n \Z i$ surveys, $\delta \tilde\mu^r = \tilde\mu^r - \langle\tilde\mu\rangle \Ns {{\rm surveys}}$. Again, we include $\delta \tilde\mu^r \in \{\delta m\Ns{B}^1, \delta x\Z 1^1, \delta c^1, ..., \delta m\Ns{B}^{n_i}, \delta x\Z 1^{n_i}, \delta c^{n_i}\}$, thus $C_{{\rm dupl}, i}^{rs}$ is a $3 {n \Z i} \times 3 {n \Z i}$ matrix. 

We also add the standard deviation of the differences in the brightness of duplicate supernovae (found by \citet{Scolnic_2022} to be $0.102$\,mag) to the off-diagonal $(m\Ns{B}^r, m\Ns{B}^s)$ matrix elements of these duplication matrices $C_{{\rm dupl}, i}$. 

Combining the duplication covariances of all $N$ unique supernovae, we obtain a block diagonal matrix
\begin{equation}
    C\Ns{dupl} = \left(\begin{array}{ccc}
        C_{{\rm dupl}, 1} & & \\
         & \ddots & \\
         & & C_{{\rm dupl}, N}
    \end{array}\right)
\end{equation}
which is of dimension $3 \mathcal{N} \times 3 \mathcal{N}$ since $\mathcal{N} = \sum_{i = 1}^{N} n \Z i$.

The full covariance matrix is then given by the sum of the statistical, systematic and duplication covariance. As all three of these matrices are constructed to be positive semi-definite, so is the full covariance matrix.

\section{Statistical Methods}\label{sec:statmethods}

Bayesian methods offer a complementary tool to frequentist methods. Given the extent to which the advantages of each approach have been debated in all scholarly enquiry, even much further back in academic publishing history than the much--cited review of \citet{Kass_1995}, it is fair to say both approaches have their uses, depending on circumstances. Bayesian methods require prior knowledge that includes models or theories as well as data. The crux of this paper, and \cref{sec:modelcomp} in particular, is just such a Bayesian comparison.

Obtaining the Bayesian evidence is computationally intensive. Therefore, a frequentist statistical approach was used to estimate the parameter values (having a faster optimisation time), and a Bayesian sampler was used to numerically compare the two models. \citet{Kass_1995} provide classifications for the strength of the evidence; with $| \ln{B} | < 1$ having no statistical preference, $1 \leq | \ln{B} | < 3$ indicating positive evidence, $3 \leq | \ln{B} | < 5$ and $5 \leq | \ln{B} |$ representing strong and very strong evidence respectively.

We follow the work of \citet{Dam_2017}, using the framework of \citet{Nielsen_2016} for the purpose of our analysis. There was a debate where \citet{Rubin_2016} criticised the methods of \citet{Nielsen_2016} for not considering Malmquist biases. However, \citet{Dam_2017} determined that the additional parameters introduced by \citet{Rubin_2016} did not significantly improve Bayesian test statistics. Therefore, for simplicity, we restrict ourselves to the framework of \citet{Nielsen_2016}.

The same priors chosen by \citet{Dam_2017} were adopted (except for a larger $\beta$ range) and are summarised in \cref{tab:priors}. The parameters can be organised in terms of three different groups, cosmological, \textit{nuisance}\footnote{Nuisance parameters are those additional to the model parameters under consideration. Many empirical parameters are in this class. However, for \sne\ analysis, $\Hn$ is also treated as a nuisance parameter as it is degenerate with $M\Ns B$ in \cref{eq:bandaid}.}, and \textit{nuisance standard deviations}.

We adopt the Bayesian hierarchical likelihood construction implemented by \citet{March_2011}, as used by \citet{Nielsen_2016}. This can be expressed in the form

\begin{align}\label{eq:likelihood}
    \mathcal{L} &\equiv \prod_{i = 1}^N \prior\left[\left.(\hat{m}\Ns{B}^{*}, \hat{x}\Z 1, \hat{c}) \Z i \right| H\right] \nonumber\\
    &=  \prod_{i = 1}^N \int \prior\left[\left.(\hat{m}\Ns{B}^{*}, \hat{x}\Z 1, \hat{c}) \Z i \right| (M\Ns{B},x\Z 1,c) \Z i, H\right] \nonumber\\
    &\hblank{8} \times \prior\left[\left.(M\Ns{B}, x\Z 1, c) \Z i \right| H\right]  \diff M\Ns{B} \diff x\Z 1 \diff c,
\end{align}
where the quantities which are denoted with a hat are considered to be observed values, and the true values (the quantities which are not denoted by a hat) are drawn from Gaussian distributions and account for fluctuations from various sources of noise \citep{Nielsen_2016, March_2011, Dam_2017}. The true data represents the intrinsic parameters utilised explicitly in the SALT2 relation.

\citet{Nielsen_2016} follow the analysis of \citet{March_2011} and adopt global, independent Gaussian distributions\footnote{For analyses not assuming Gaussian distributions see \citep{Seifert_2024}.} for $M \Ns B$, $x\Z 1$ and $c$ to determine the probability density of the true parameters \citep{Nielsen_2016, March_2011}. This is determined by $\prior[(M \Ns B,x\Z 1,c)|\Theta] = \prior(M \Ns B|\Theta) \times \prior(x\Z 1 | \Theta) \times \prior(c | \Theta)$; where for $\tilde\mu \in \{M \Ns B, x\Z 1, c\}$, the probability can be expressed as $\prior(\tilde\mu|\Theta) \equiv (2 \pi \sigma_{\tilde\mu}^2)^{-1/2} \times \exp\left(-\frac12\left[(\tilde\mu-\tilde\mu_0)/\sigma_{\tilde\mu_0}\right]^2\right)$. By creating a vector of the true parameters, $Y=[M\Ns{B,1}, x\Z{1,1}, c\Z1, ...,M\Ns{B,\mathcal{N}}, x\Z{1,\mathcal{N}}, c\Z{\cal N}]^\intercal$ and a diagonal matrix of dispersion uncertainties $\Sigma \Z {l} = \mathrm{diag}(\sigma_{M\Z0}^2,\sigma_{x\Z{1,0}}^2,\sigma_{c\Z0}^2,...)$, this can be constructed by 
\begin{equation}\label{eq:Ptrue}
    \prior(Y|\Theta) \equiv \det\left(2 \pi \Sigma \Z {l}\right)^{-1/2} \exp\left(-\frac{1}{2}\left(Y-Y\Z0\right)^\intercal\Sigma \Z {l}^{-1}\left(Y-Y\Z0\right)\right),
\end{equation}
where $Y\Z0 = [M\Ns{B,0}, x\Z{1,0}, c\Z0, M\Ns{B,1}, ...]^\intercal$ is the zero-point vector of the parameters, drawn from the intrinsic Gaussian averages.

To transform the absolute magnitude to an apparent magnitude minus the distance moduli, the block diagonal matrix $A$ given by
\begin{equation}\label{eq:AMat}
    A = \left(\begin{array}{cccc}
        1 & -\alpha & \beta &  \\
        0 & 1 & 0 & 0 \\
        0 & 0 & 1 &  \\
         & 0 &  & \ddots \\
    \end{array}\right),
\end{equation}
is introduced, which transforms as $Y\Z 0 A = [M\Ns{B,0} - \alpha x\Z{1,0} + \beta c\Z 0, x\Z{1,0}, c\Z 0, M\Ns{B,1} - \alpha x\Z{1,1} + \beta c\Z 1, ...]^\intercal$. Furthermore, the estimated experimental covariance matrix of statistical and systematic errors, $\Sigma_{d}$, and the new vector $\hat{Z}_{\Delta\mu\Z 0}$ are considered. $\hat{Z}_{\Delta\mu\Z 0}$ is constructed from the observational values for $m\Ns{B}^*$, $x\Z 1$ and $c$ by $\hat{Z}_{\Delta\mu\Z 0}=[\hat{m}\Ns{B,1}^{*} - \mu\Z 1 - \Delta\mu\Z 0, \hat{x}\Z {1,1}, \hat{c}\Z 1, ...,\hat{m}\Ns{B,\mathcal{N}}^{*} - \mu\Ns {\cal N} - \Delta\mu\Z 0, \hat{x}\Ns {1,\mathcal{N}}, \hat{c}\Ns {\cal N}]^\intercal$ and transforms as $YA$. Together, this yields the likelihood
\begin{align}
    \mathcal{L}_{\Delta\mu_0} &= \int \prior(\hat{Z}_{\Delta\mu_0} | Y, \Theta) \ \prior(Y | \Theta) \ \diff Y\nonumber\\
    &= \det\left[2 \pi \left(\Sigma \Z d + A \Sigma \Z l A^\intercal \right)\right]^{-1/2} \nonumber\\
    &\hblank{3}\times \exp\left[-\frac{1}{2}\left(\hat{Z}_{\Delta\mu\Z 0}-Y_0 A\right)^\intercal\left(\Sigma \Z {d} + A \Sigma  \Z {l} A^\intercal\right)^{-1} \left(\hat{Z}_{\Delta\mu\Z 0}-Y_0 A\right)\right]\label{eq:likelihoodFull},
\end{align}
for given zero-point offset $\Delta\mu\Z 0$. 

This parameter must be included to account for the possibility that the Hubble constant varies between subsamples as different redshift cuts are made. In standard analyses where the whole sample is considered without implementing a sequence of redshift cuts, this question is only considered once in a single marginalisation step. The offset indicates a universal degeneracy between the choice of the Hubble constant and the absolute magnitude of the supernovae. Naturally, this {``zero-point calibration''} issue is intimately connected to the Hubble tension. Our aim here is to compare two cosmological models independently of the zero-point issue, using Bayesian methods. Consequently, while the zero-point issue is of considerable interest, it is temporarily set aside in making redshift cuts by marginalising over $\Delta\mu\Z 0$. A similar approach was adopted in past analyses of the subsamples \citep{Gaur_2019,Lane_2022} of the Pantheon and Pantheon+ surveys common to JLA, in which only distance moduli fit with FLRW geometry assumptions were available.
Considering $\hat{Z}_{\Delta\mu\Z 0}-Y_0 A$ and thus $\mathcal{L}_{\Delta\mu\Z 0}$ as a function of $y + \Delta \mu\Z 0$, the 
reduced likelihood obtained by the marginalisation is given by
\begin{align}
    \mathcal{L} &= \frac{1}{6 \sigma \Z y} \int_{- 3 \sigma \Z y}^{3 \sigma \Z y} \mathcal{L}_{\Delta\mu\Z 0}(y + \Delta \mu\Z 0) \diff \Delta \mu\Z 0\label{eq:likelihoodMarg},
\end{align}
where $y = M\Ns{B,0} - \alpha x\Z{1,0} + \beta c\Z 0$ is the zero-point value of $m\Ns{B}^* - \mu$ and $\sigma_y = \sqrt{\sigma_{M_0}^2 + \alpha^2\sigma_{x\Z{1,0}}^2 + \beta^2\sigma_{c_0}^2}$ is its standard deviation.
As the marginalisation slows down the convergence significantly, it is performed for the frequentist analysis only. We analyse its effects for this approach but calculate the Bayesian results based on the original likelihood from \cref{eq:likelihoodFull}.

The \citet{Nielsen_2016} approach, however, has some flaws in its formulation, that are expanded upon and corrected in the work of \citet{Seifert_2024}. \citet{March_2011} and \citet{Nielsen_2016} assumed that there is no correlation between the true parameters $M\Ns B$, $x\Z1$, and $c$, which is not true. Factors such as the host environment's metallicity and circumstellar medium influence both the stretch and colour, making them correlated and inseparable. We can assume that because $M\Ns B$ is a constant (and by definition of having zero stretch and colour) there is no true correlation between $M\Ns B$ and $x\Z1$ or $c$. The next assumption that \citet{March_2011} and \citet{Nielsen_2016} make is that the $x\Z1$ and $c$ parameters follow Gaussian distributions. While this was a reasonable assumption at the time as there was believed to be a systematic, it is now believed that the distribution follows an asymmetric (skewed) Gaussian. The assumption of imposing Gaussianity may emphasise different characteristics of the distribution. A comparison between the asymmetric correlated Gaussian, and a Tripp parameter version can be found in \citet{Seifert_2024}. However, in the present analysis, we restrict to the approach by \citet{Nielsen_2016} and \citet{March_2011} to be able to compare to previous work.

We utilised a nested Bayesian sampler \texttt{PyMultiNest} \citep{Buchner_2014}, which interacts with the \texttt{MultiNest} \citep{Feroz_2008, Feroz_2009, Feroz_2019} code. This was used to compute the evidence and determine the Bayes factor in comparing the \lcdm\ and timescape models. The accuracy of \texttt{PyMultiNest} is determined by the number of `live' points used in the sampling, with the error of order $\sigma \sim \mathcal{O}(n\Ns{live}^{-1/2})$; and the tolerance for the nested sampling. We chose a tolerance of $10^{-3}$ and $n\Ns{live} = 1000$ for nine parameters. 

\section{Results}\label{sec:results}

Following \citet{Dam_2017}, we perform a likelihood analysis for subsets of the P+1690 sample determined by redshift cuts $z\Ns{CMB} \geq \zmin$ for varying $\zmin$ (\cref{fig:random}). We investigate the light-curve parameters obtained in this way and compare the results for analyses started from different subsamples instead of the full P+1690 sample. 

Since the light-curve parameters exhibit variations across different redshift cuts, we will refine the concept of the Scale of Statistical Homogeneity (SHS). We first denote SHS measures derived empirically from $n$--point galaxy correlation functions as the {\em SHS$_{{\rm g}n}$ statistics}. Thus conventional analyses \citep{Hogg_2005, Sylos_Labini_2009, Scrimgeour_2012} based on 2-point galaxy correlation functions investigate the SHS$_{{\rm g}2}$ statistic using {a $\Lambda$CDM cosmology}. Our analysis here will not be based on higher order galaxy correlation functions, but on the empirical quantities $\beta c$ (colour) and $\alpha x\Z 1$ (stretch) that appear in the Tripp relation \cref{eq:tripp}.
We, therefore, introduce the statistics SHS$_\alpha$ (stretch) and SHS$_\beta$ (colour) to explore this phenomenon in \cref{sec:shs}. The SHS$_\alpha$ statistic, derived from the light-curve parameters, serves as an indicator of the point at which the results of \sne, free from biases and subject to numerical uncertainty, approach a statistically homogeneous Universe. This is a data-driven definition of the SHS and should be model-independent, independent even from the FLRW metric commonly used. The analysis of the light-curve parameters, the random subsamples and the SHS$_\alpha$ are based on frequentist methods, involving marginalised likelihoods if needed. However, for the evaluation and comparison of the two models, we consider both the frequentist and Bayesian approach as described in \cref{sec:statmethods}, and presented in \cref{sec:modelcomp}.

\subsection{Pantheon+/JLA Common Subsample} \label{sec:commonsubsample}

\begin{figure}
	\includegraphics[width=\columnwidth]{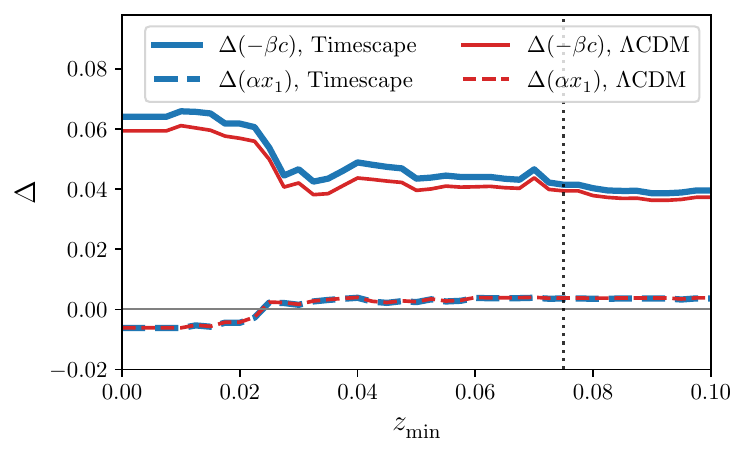}
    \caption{The differences ($\Delta$) between the $\alpha x\Z 1$ and $-\beta c$ Tripp equation terms calculated for the \pplus\ and JLA common subsample (\cref{eq:tripp}). The differences between the two datasets asymptotically approach $0.00360 \pm 0.00013$ ($0.00378 \pm 0.00012$) for $\alpha x\Z 1$ and $0.0405 \pm 0.0022$ ($0.0381 \pm 0.0021$) for $- \beta c$ based on the timescape (\lcdm) model. These differences arise from the change from SALT2 for the JLA analysis to SALT2-2021 for \pplus\ \citep{Taylor_2021}. The vertical dotted line represents the statistical homogeneity scale (SHS$_\alpha$) we define to be $z\Ns{CMB}=0.075$.}
    \label{fig:difference_subsamples_JLA}
\end{figure}

To compare the \pplus\ results to the ones by \citet{Dam_2017} based on the JLA catalogue, we compute the common subsample of the two supernovae catalogues. This common subset consists of 584 \sne, for which both the JLA and \pplus\ data are accessible. Within this set of 584 supernovae, it was necessary to exclude four of them in order to ensure that the covariance matrix remains positive semi-definite (as detailed in \cref{sec:statcov} and \cref{app:statdata}). As a result, we will refer to this sample as \textit{P+580}.

Due to the presence of duplicate supernova events, P+580 comprises 635 supernova observations. For the sake of direct comparison, we focus on the JLA data encompassing the same 580 unique supernovae as those found in P+580, denoting this dataset as \textit{JLA580}. Our analysis encompasses both datasets, revealing significant disparities in the light-curve parameters. Notably, the asymptotic difference in the $-\beta c$ term, as described by \cref{eq:tripp}, amounts to $0.0381 \pm 0.0021$ when assuming a \lcdm\ cosmology (as shown in \cref{fig:difference_subsamples_JLA}). This difference can be attributed to the alterations in the SALT2 surface, as documented by \citet{Taylor_2021}.

\subsection{Low-Redshift Supernovae} \label{sec:reducelowz}

\begin{figure}
	\includegraphics[width=\columnwidth]{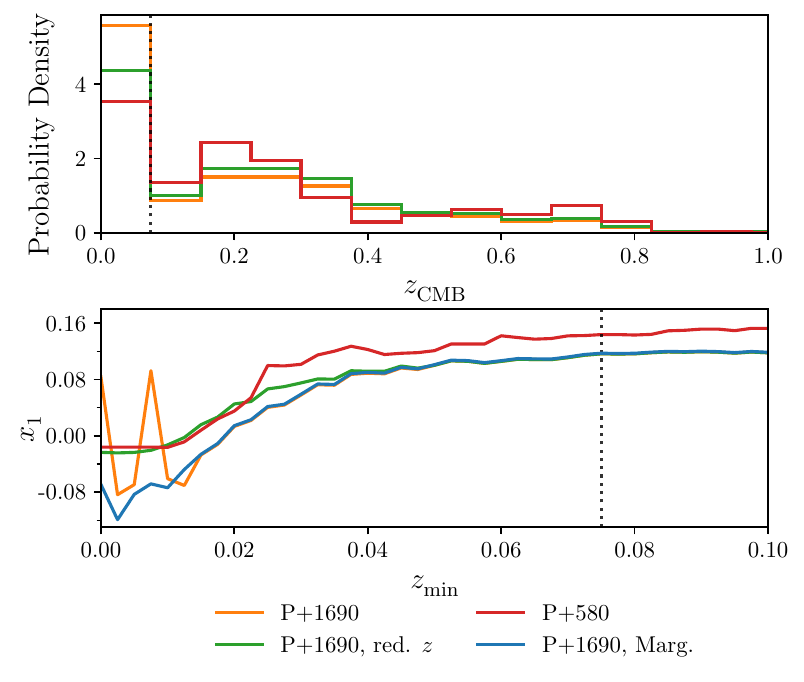}
    \caption{The distribution of the \sne\ redshift in the CMB frame (upper) and its influence on the $x\Z 1$ parameter for the subsamples (lower, for \lcdm\ model). The spikes in the low redshift regime for the P+1690 sample can be avoided by choosing a reduced low-redshift sample or using the marginalising approach. The vertical dotted line represents the SHS$_\alpha$ found at $z\Ns{CMB}=0.075$.}
    \label{fig:lesslowz}
\end{figure}

The light-curve parameters (e.g., the stretch $x\Z 1$, shown in \cref{fig:lesslowz}) have a large scatter for low redshift cuts when fitted to the full \pplus\ dataset. This behaviour is not observed when analysing the full JLA sample \citep{Dam_2017}, nor does it appear when considering the P+580 subsample. To determine if this feature is redshift dependent, we construct a new subsample \textit{P+1690, reduced z}, which mimics the P+580 distribution of supernovae for $z\Ns{CMB} < 0.05$. As this dataset has minimal scatter (\cref{fig:lesslowz}), we confirm that this artefact is due to the high number of low redshift supernovae with a wide range of $x\Z 1$ and $c$ parameters reported by \pplus. As discussed in \cref{sec:statmethods}, different subsamples of supernova observations within the full dataset might rely on different values for the Hubble constant due to varying redshift cuts. This is a known source for variations throughout the full sample and is accounted for by marginalisation over the zero-point offset by choosing a different likelihood (\cref{eq:likelihoodMarg}). As shown in \cref{fig:lesslowz}, this approach also resolves the artefacts in the low redshift regime.

\subsection{Random Subsamples} \label{sec:random}

\begin{figure*}
    \includegraphics[width=\textwidth]{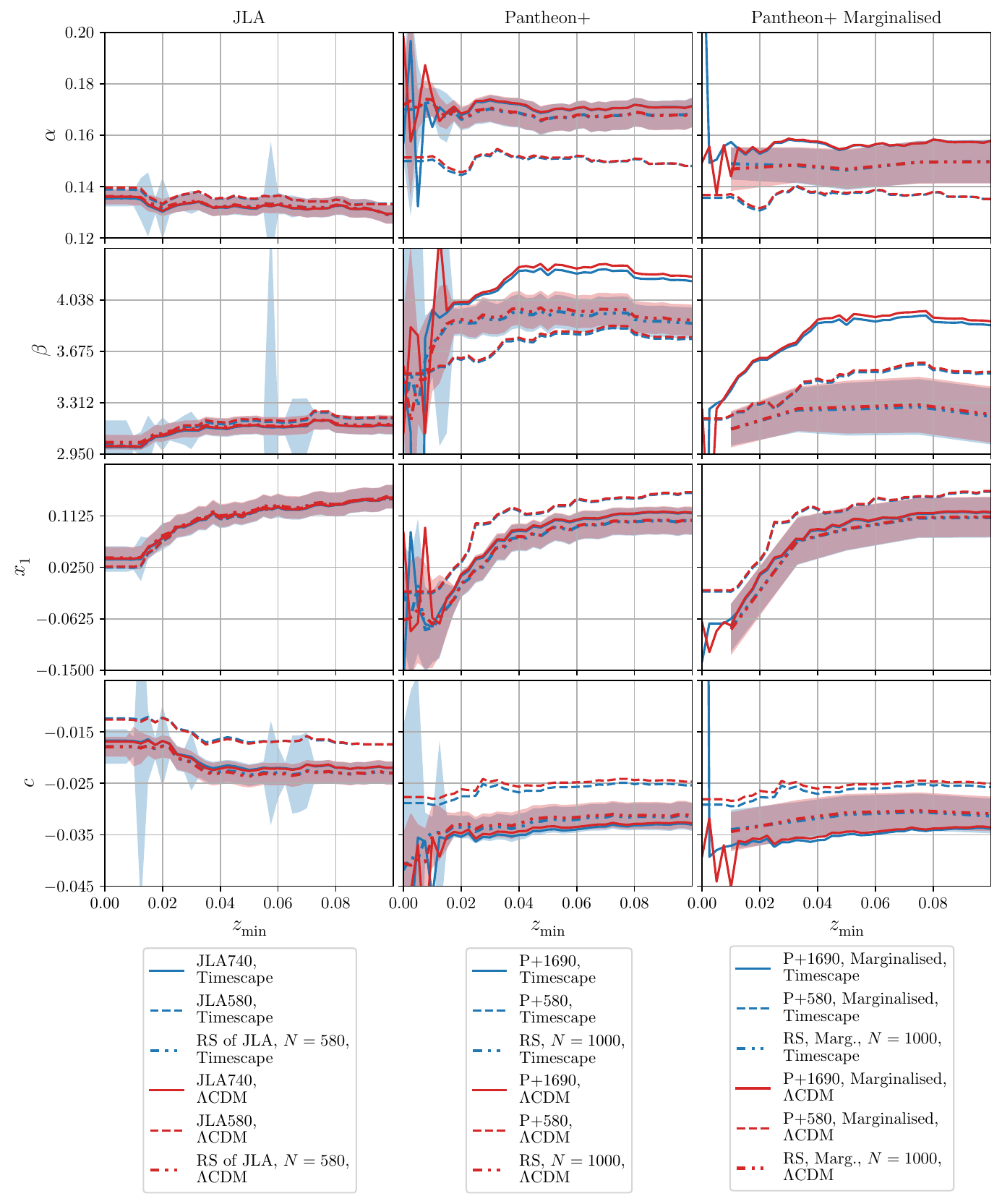}
    \caption{The SALT2 parameters for JLA (left), \pplus\ (middle) and the marginalised results (right). The shaded regions give one standard deviation around the median of the parameters at a given $\zmin$ calculated from our set of random subsamples (referred to as \textit{RS}). The right column shows results which have been integrated with \cref{eq:likelihoodMarg} to account for the choice of Hubble constant. For the offset between \pplus\ data (middle column) and JLA data (left column) see \cref{fig:difference_subsamples_JLA}. For plots of the products $\alpha x\Z 1$ and $-\beta c$ that contribute to the distance modulus $\mu$, see \cref{fig:lightcurveparams_products}.}
    \label{fig:random}
\end{figure*}

The asymptotic results for the light-curve parameters are not numerically stable, but instead, depend on the selected subsample of the \pplus\ dataset. $50$ random subsamples were chosen of size $N = 1000$, where $N$ refers to the number of unique supernovae chosen (the actual random subsamples may include more individual supernova observations). The median and standard deviation of the light-curve parameters for varying $\zmin$ computed from these samples are shown in \cref{fig:random} (middle column) alongside the full sample and the \common. \citet{Wagner_2019} use a similar method of taking random subsamples to systematically investigate the sampling issues in the Pantheon catalogue.

We conduct a similar analysis with the JLA data, choosing the size $N$ of the random samples to match that of the common subsample ($N = 580$). The results (left column of \cref{fig:random}) show that the full sample is within $2\sigma$ of the randomised results. 

For the \pplus\ common dataset (P+580), we observe important differences when comparing it to the full sample and the random subsamples of $N=1000$ (middle column of \cref{fig:random}). The values of $\alpha$, $\beta$, $x\Z 1$ and $c$ are within the 2--3$\sigma$ bounds of the full sample, and within 1--2$\sigma$ of P+580. These variations required further analysis to identify whether the value changes are a product of taking random subsamples of differing sizes.

\begin{figure}
	\includegraphics[width=\columnwidth]{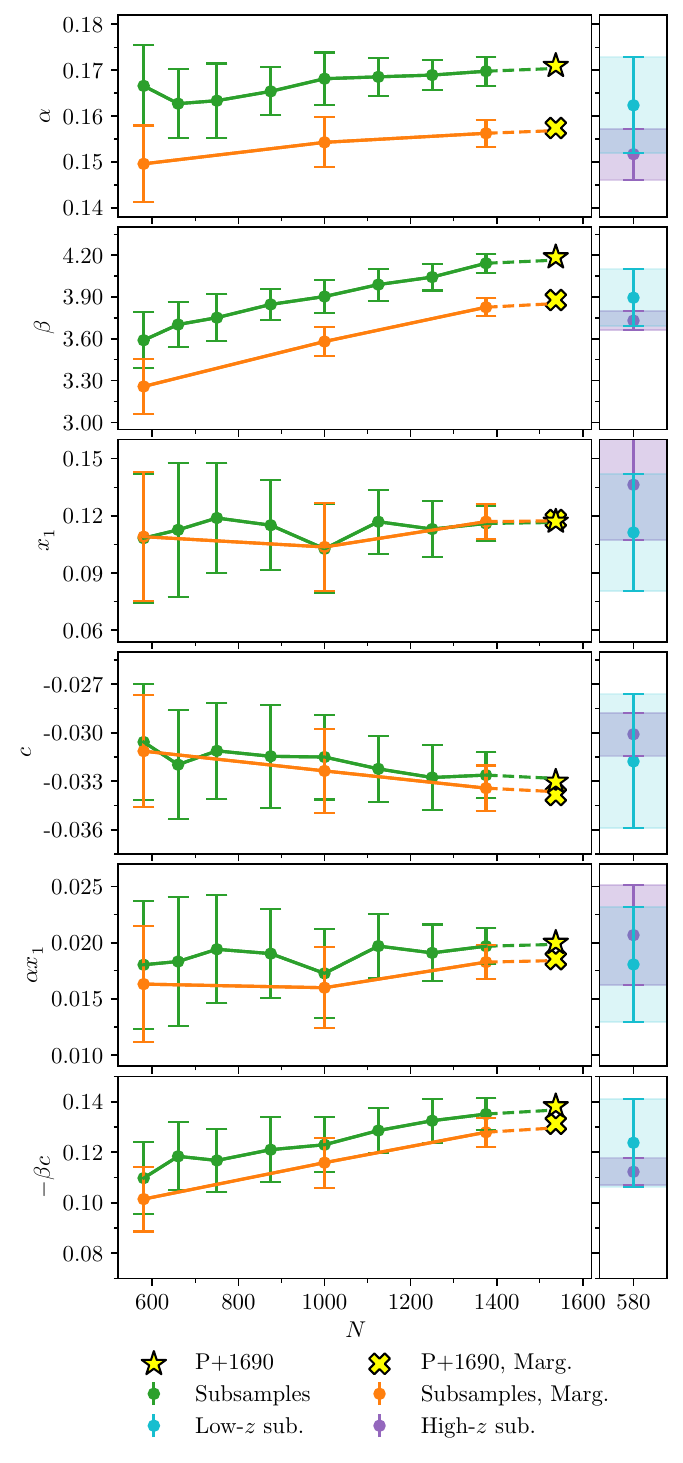}
    \caption{The change in the light-curve parameters for increasing number of unique \sne\ ($N$) for the random samples for the timescape model (left) and tests of the effects of distribution on the sample (right). The medians of the samples were taken of all values after the SHS$_\alpha$ ($z\Ns{CMB} = 0.075$ up to $\zmin = 0.1$) for both the left (50 random subsamples) and right columns (containing 10 different random subsamples). The star and cross markers indicate the values of the full (non-marginalised) P+1690 and marginalised P+1690 evaluations respectively. Note that the P+1690 sample consists of $N = 1535$ unique \sne. The dashed lines indicate the prospective advancement with sample size.}
    \label{fig:conv}
\end{figure}

As the common subsample is much smaller than $N = 1000$, we investigate the possible effects of the sample size on the scatter of random subsamples. In particular, we calculated the light-curve parameters of subsamples of P+1690 for\footnote{The lowest value for $N$ was chosen for comparison with P+580.} $N = 580,\,660,\,750,\,875,\,1000,\,1125,\,1250,\,1375$ at\footnote{We choose $\zmin = 0.033$ as an estimate for the SHS$_\alpha$ that \citet{Dam_2017} used for their results.} $\zmin = 0.01,\,0.033,\,0.05,\,0.075,\,0.1$. Up to statistical fluctuations, all of these random subsamples have the same redshift distribution as the full P+1690 sample. 

The model parameters, which vary with sample size ($N$), are depicted in \cref{fig:conv}. Our analysis reveals that for the non-marginalised calculations using the likelihood from \cref{eq:likelihoodFull}, the parameter $\beta$ demonstrates a clear dependence on the sample size, whereas $\alpha$, $x\Z 1$, $c$, and $\Omega\Ns{M0}$ fall comfortably within the uncertainty bounds of the entire sample.

$\beta$ converges in parameter-space for each of the various samples across differing redshift cuts; however, these parameters do not converge to a uniform value for all sample sizes as expected. Interestingly, increasing the size of each sample appears to increase the $\beta$ value for both the marginalised and non-marginalised samples, therefore this phenomenon is not attributable to a specific choice of the Hubble parameter.

To investigate what influence the redshift distribution has on the $\beta$ convergence, we compared various samples of size $N = 580$ with differing redshift distributions. For the samples highlighted in green in \cref{fig:conv}, we selected all \sne\ with an equal probability, thereby resulting in a redshift distribution nearly identical to that of the full sample. For the `high' sample, we considered the 290 \sne\ from the P+1690 dataset with the highest redshift (or with the lowest redshift for the `low' sample) and selected the remaining 290 \sne\ from the remainder of the P+1690 dataset using the same methodology as the uniform sampling. 

As displayed in \cref{fig:conv}, the `low' sample is consistent with the full sample for all parameters, including $\beta$. This confirms that the behaviour observed for the uniform subsamples is caused by the redshift distribution. For further analysis of the parameter evolution, a full sample with a uniform distribution would be needed.

\begin{figure}
    \includegraphics[width=\columnwidth]{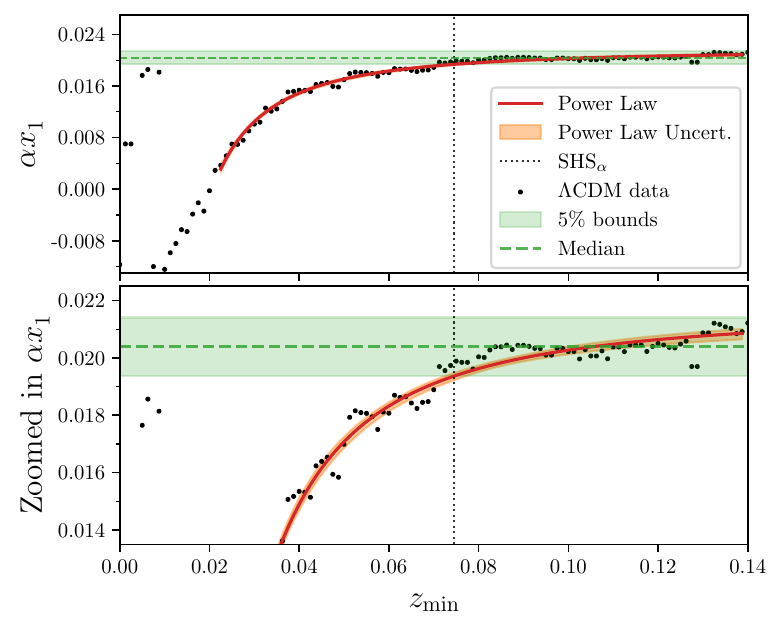}
    \caption{The convergence of the $\alpha x\Z 1$ light-curve parameter for the \lcdm\ model across various redshift cuts. In the upper plot, a power--law model has been fit to the data, and the green shaded region represents within 5\% of the median value within the range $0.1 \leq \zmin < 0.14$ indicating when the model converges. The vertical dotted line represents the SHS$_\alpha$ found at $\zmin = 0.075$. The lower plot provides an enlarged view of the data to illustrate finer details.}
    \label{fig:shs}
\end{figure}

\begin{figure}
	\includegraphics[width=\columnwidth]{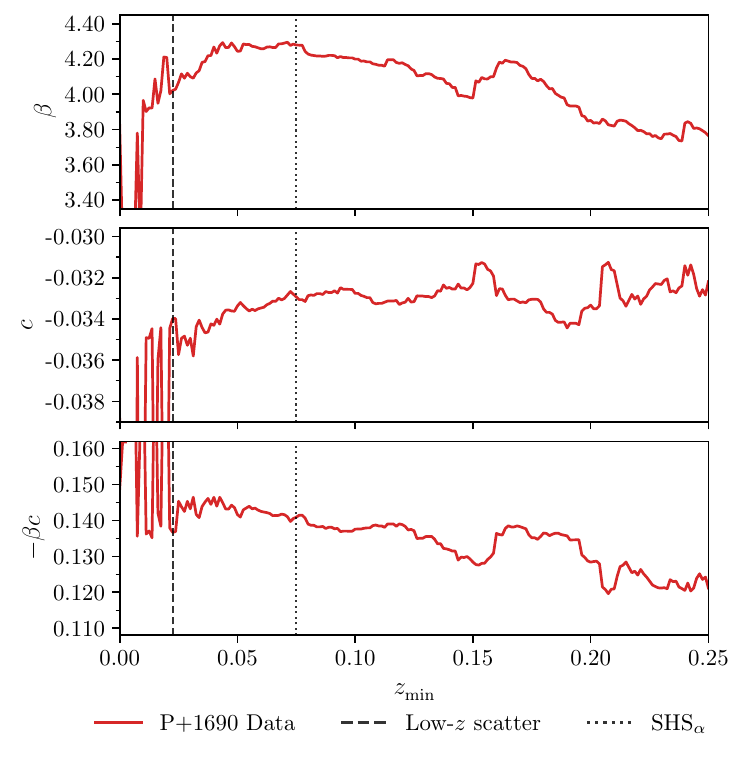}
    \caption{Light-curve parameters $\beta$ and $c$ and their degenerate product for the \lcdm\ model at higher redshifts to analyse how they evolve with redshift cuts. The low-redshift scatter shows the power--law model lower bound for our SHS$_\alpha$ analysis. Here we see the effects of Malmquist bias on our parameters, making us unable to use them for determining an SHS$_\beta$ as an alternative to the SHS$_\alpha$. For the analysis of $\alpha$, $x\Z 1$ and $\alpha x\Z 1$, see \cref{fig:ext_z_x1}.}
    \label{fig:ext_z_c}
\end{figure}

\subsection{Scale of Statistical Homogeneity}\label{sec:shs}

\citet{Dam_2017} report significant changes in the light-curve parameters in the presence of an emerging SHS. For the JLA sample, a change in the parameters $\beta$, $x\Z 1$ and $c$ across the differing redshift cuts is clearly visible in \citet[Fig.~2]{Dam_2017} and in the corresponding panels of \cref{fig:random}. Specifically, the analysis of randomly subsampled data validates the conclusions drawn from the entire JLA dataset. 

Our definition of the SHS$_\alpha$ is exclusively rooted in empirical light-curve fitting parameters. This approach is adopted to ensure the independence of cosmological models and eliminate any assumptions associated with the perturbed FLRW geometry. We achieved this by fitting a power--law model to the $\alpha x\Z 1$ parameter, represented by $-az^{-n} + C$, where $a$, $n$, $C$ are parameters subject to fitting\footnote{These parameters were fit with the Markov Chain Monte-Carlo sampling tool \texttt{emcee} \citep{Foreman-Mackey_2013} with values of $a = (23.2 \pm 3.5) \times 10^{-6}$, $n = 1.76 \pm 0.04$, and $C = (21.61 \pm .10) \times 10^{-3}$.}, and $z$ represents the redshift cut. 

\begin{figure}
    \centering
    \includegraphics[width=\columnwidth]{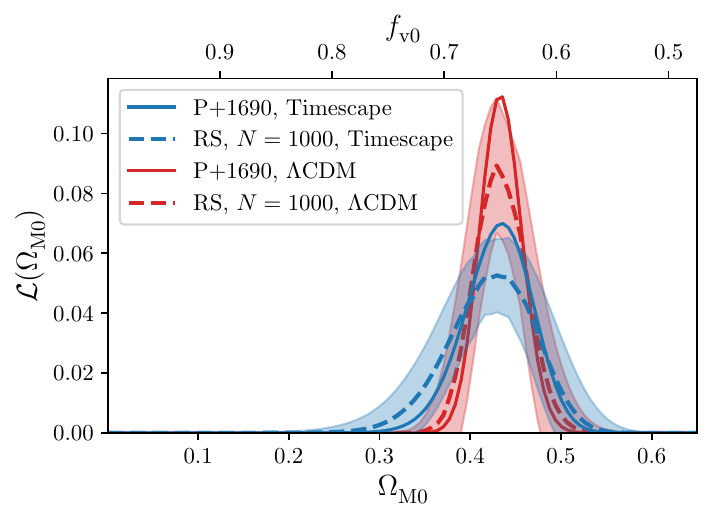}
   \caption{Normalised likelihood at $z\Ns{CMB} = 0.033$ of the model parameters, $\Omega\Ns{M0}$ for the \lcdm\ and $f\Ns{ v0}$ for the timescape model for \pplus\ and for the 50 $N=1000$ random subsamples (RS). The maximum likelihood values are $f\Ns{v0, MLE} = 0.675$, and $\Omega\Ns{M0, MLE} = 0.433$, respectively, for the P+1690 sample. The integrated non-normalised likelihood of the P+1690 sample is $2.73\times 10^{167}$ for \lcdm\ and $2.06\times 10^{167}$ for timescape.}
    \label{fig:profilelike}
\end{figure}

\begin{figure}
    \includegraphics[width=\columnwidth]{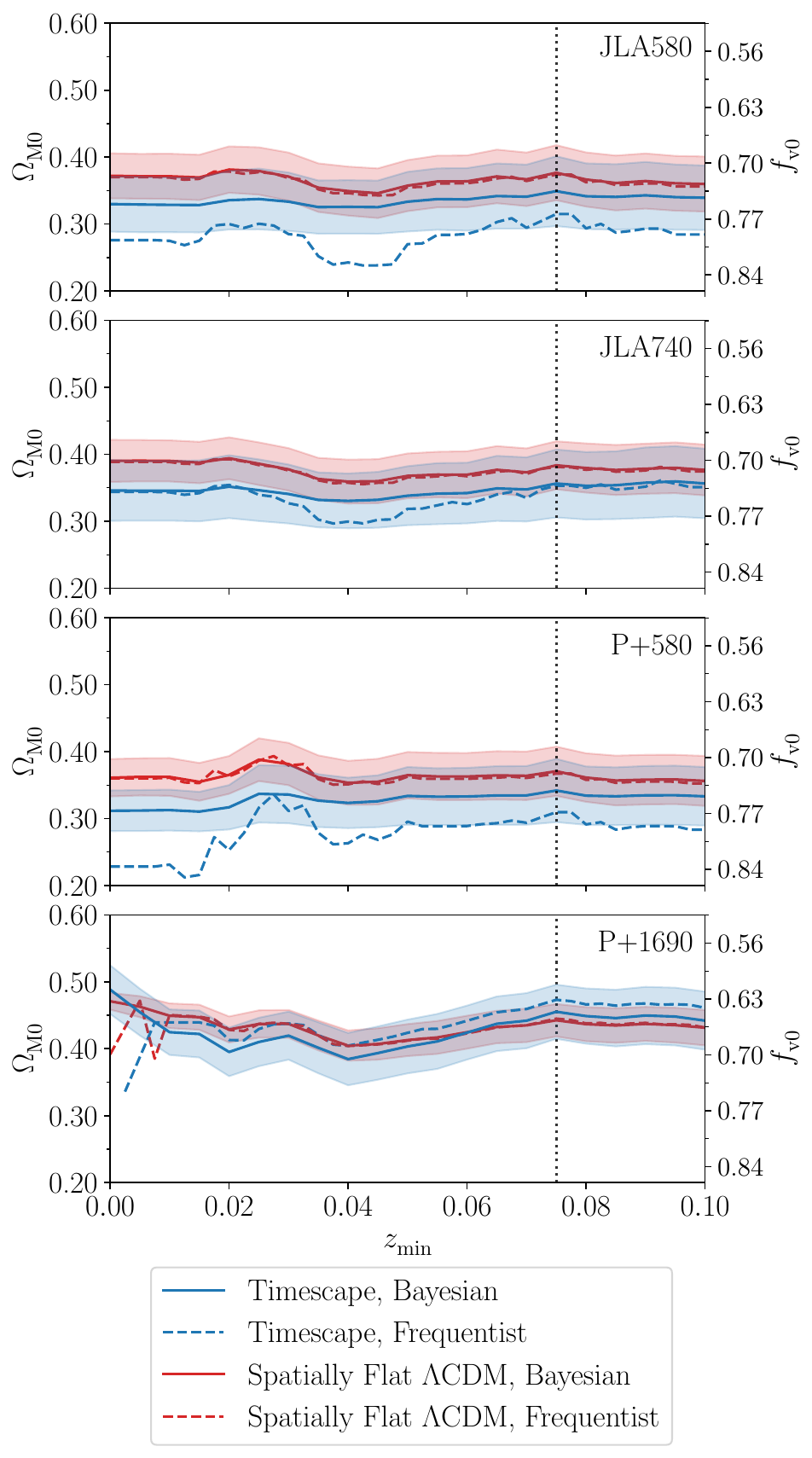}
    \caption{Model parameter $\Omega\Ns{M0}$ for \lcdm\ and $f\Ns{v0}$ for the timescape model, evaluated with Bayesian and frequentist statistical methods. The dotted line represents the SHS$_\alpha$.}
    \label{fig:omega}
\end{figure}

We consider the redshift cut to be within the SHS$_\alpha$ when the power--law falls within $5\%$ of the limit of $\alpha x\Z 1$ (with the uncertainties coming from 3 and 7\% intervals). This limit is estimated by the median value of $\alpha x\Z 1$ in the range $0.10 \leq \zmin < 0.14$, which we chose as to when the parameter has obviously converged, as seen in \cref{fig:shs}. 

We employed the power--law model within the range of $0.0225 \leq \zmin < 0.14$, where the bounds were chosen such that they avoid low-redshift noise and reduced SNe sampling at higher redshift (for additional details, refer to \cref{sec:reducelowz}). From this fit we find a parameter convergent SHS$_\alpha$ of $z\Ns{CMB} = 0.075^{+0.007}_{-0.009}$, which is significantly greater than values estimated in earlier studies \citep{Dam_2017, Hogg_2005, Scrimgeour_2012}. However, as opposed to the statistical homogeneity scale SHS$_{\rm g 2}$ considered by \citep{Hogg_2005,Scrimgeour_2012}, our definition of the SHS$_\alpha$ does not use the two-point galaxy correlation function but relies on the convergence of the light-curve parameters and the distance modulus. This approach respects general features in the light-curve parameters arising from unaccounted large scale structures that are only now apparent using the \pplus\ dataset (further discussed in \cref{sec:modelcomp}). It thus leads to a higher value of the SHS$_\alpha$.

To check the impact of the high redshift cut on the fit, we extend the upper bound of $\zmin$ to be 0.25. Parameter values at $\zmin = 0.25$ have nearly 20\% less unique \sne\ than $\zmin = 0.14$ and therefore are more prone to numerical fluctuations with increasing redshift cut. For the extended range we find a SHS$_\alpha$ of $z\Ns{CMB} = 0.086^{+0.009}_{-0.013}$, which is within 1$\sigma$ of our stated value. We therefore use our initial fit over the range $0.0225 \leq \zmin < 0.14$ for the SHS$_\alpha$.

To conduct our convergence analysis, we focus on the $\alpha x\Z 1$ parameter to mitigate numerous potential issues stemming from Malmquist bias, which falls beyond the scope of this current analysis. When examining the $\beta$ or $c$ parameters there is no convergence within the tested range, thus a SHS$_\beta$ can not be defined based on the P+1690 sample.

The $\beta$ `constant' parameter exhibits a pattern of fluctuation, peaking at a value exceeding 4.30 just after $z\Ns{\rm min} = 0.07$ and subsequently decreasing to 3.77 at $z\Ns{\rm min} = 0.25$ (as shown in \cref{fig:ext_z_c}). However, this trend is counterbalanced by the supernovae appearing `bluer' as we extend our observations to higher redshifts. When we investigate the degeneracy between $\beta$ and the colour parameter $c$, we are able to discern and elucidate the influences of Malmquist bias on our data as we progress through different redshift cuts (\cref{fig:ext_z_c}).

We also did not solely utilise the $\alpha$ parameter for our analysis. This was to be in alignment with previous studies (e.g., \citet{Dam_2017}), which reported minimal variations in its value within the region of an emerging SHS$_\alpha$. Furthermore, it exhibits a significant decline at $z\Ns {min} \sim 0.17$, rendering it an unreliable indicator for recovering the SHS$_\alpha$ (see  \cref{fig:ext_z_x1}). Nevertheless, it does play a valuable role in constraining the scatter associated with the lower bound.

The decrease in the $\alpha$ parameter, however, is effectively balanced by an increase in $x\Z 1$, resulting in the $\alpha x\Z 1$ degenerate parameter remaining nearly constant within the tested high-redshift range (\cref{fig:shs} and \cref{fig:ext_z_x1}).

We explored various convergence methods, including statistical analyses of the parameter spaces and gradients based on high-redshift cuts. However, the {power--law methodology of \cref{sec:shs}} demonstrated the advantage of requiring minimal human intervention, thereby allowing for a more standardised approach to the analysis.

More tests are required to determine if the 3.9$\sigma$ difference between the SHS$_{{\rm g}n}$ and SHS$_\alpha$ is due to astrophysical modelling systematics, or the choice of the \LCDM\ cosmological model used in the SHS$_{{\rm g}n}$ methodology. While the SHS$_\alpha$ analysis includes the current appropriate systematic effects for both the timescape and $\Lambda$CDM cosmological models (c.f. \cref{sec:systcov}), the inclusion of other systematics such as Malmquist bias and the inclusion of peculiar velocities may alleviate some of the tension between the two methods. However, it is plausible that the reason for the tension is related to unknown or undeveloped \sne\ and/or galaxy-correlation systematics. 

It is worth mentioning that the higher SHS value does not directly conflict with the \LCDM\ universe. Instead, it could be attributed to the greater calibration and sensitivity of the \sne\ sample, allowing for the impacts of the finer structure at closer redshifts to be discerned. This heightened sensitivity may result in the regime below the {SHS, for which FLRW models have reduced effectiveness,} being larger than initially anticipated from \LCDM\ model predictions using the appropriate known astrophysical systematics.

\begin{table*}
\centering
\caption{Median and standard deviation of the model and light-curve parameters beyond the scale of statistical homogeneity, $z = 0.075$, for the different samples, models and statistical methods.}
\begin{tabular}{cccccccc}
\toprule
       &              &      &    $\Omega_{\rm M0}$ &             $\alpha$ &                $x\Z 1$ &              $\beta$ &                     $c$ \\
\hline \hline
JLA580 & TS & Freq. &    0.293 $\pm$ 0.011 &  0.1355 $\pm$ 0.0008 &    0.146 $\pm$ 0.004 &    3.192 $\pm$ 0.015 &  -0.01849 $\pm$ 0.00029 \\
       &              & Bay. &    0.305 $\pm$ 0.018 &  0.1358 $\pm$ 0.0010 &    0.148 $\pm$ 0.009 &    3.188 $\pm$ 0.023 &    -0.0176 $\pm$ 0.0008 \\
       & $\Lambda$CDM & Freq. &    0.361 $\pm$ 0.006 &  0.1354 $\pm$ 0.0009 &    0.146 $\pm$ 0.004 &    3.197 $\pm$ 0.016 &    -0.0184 $\pm$ 0.0003 \\
       &              & Bay. &    0.362 $\pm$ 0.007 &  0.1365 $\pm$ 0.0009 &    0.153 $\pm$ 0.009 &    3.208 $\pm$ 0.012 &    -0.0182 $\pm$ 0.0008 \\ \hline
JLA740 & TS & Freq. &    0.352 $\pm$ 0.004 &  0.1314 $\pm$ 0.0010 &    0.137 $\pm$ 0.005 &    3.151 $\pm$ 0.015 &  -0.02198 $\pm$ 0.00020 \\
       &              & Bay. &    0.338 $\pm$ 0.009 &  0.1315 $\pm$ 0.0013 &    0.135 $\pm$ 0.004 &    3.147 $\pm$ 0.028 &    -0.0219 $\pm$ 0.0007 \\
       & $\Lambda$CDM & Freq. &  0.3769 $\pm$ 0.0027 &  0.1314 $\pm$ 0.0011 &    0.138 $\pm$ 0.005 &    3.154 $\pm$ 0.015 &  -0.02191 $\pm$ 0.00023 \\
       &              & Bay. &    0.373 $\pm$ 0.007 &  0.1319 $\pm$ 0.0015 &    0.137 $\pm$ 0.003 &    3.162 $\pm$ 0.022 &    -0.0219 $\pm$ 0.0008 \\ \hline
P+580 & TS & Freq. &    0.289 $\pm$ 0.009 &  0.1490 $\pm$ 0.0009 &    0.148 $\pm$ 0.004 &    3.765 $\pm$ 0.028 &  -0.02516 $\pm$ 0.00016 \\
       &              & Bay. &    0.291 $\pm$ 0.008 &   0.150 $\pm$ 0.0009 &    0.147 $\pm$ 0.007 &      3.76 $\pm$ 0.03 &    -0.0254 $\pm$ 0.0009 \\
       & $\Lambda$CDM & Freq. &    0.355 $\pm$ 0.005 &  0.1491 $\pm$ 0.0009 &    0.150 $\pm$ 0.004 &    3.778 $\pm$ 0.029 &  -0.02439 $\pm$ 0.00018 \\
       &              & Bay. &    0.359 $\pm$ 0.007 &  0.1489 $\pm$ 0.0009 &    0.153 $\pm$ 0.005 &      3.79 $\pm$ 0.03 &    -0.0244 $\pm$ 0.0014 \\ \hline
P+1690 & TS & Freq. &    0.467 $\pm$ 0.003 &  0.1710 $\pm$ 0.0004 &  0.1172 $\pm$ 0.0011 &    4.185 $\pm$ 0.025 &  -0.03304 $\pm$ 0.00016 \\
       &              & Bay. &    0.462 $\pm$ 0.009 &  0.1688 $\pm$ 0.0013 &    0.119 $\pm$ 0.005 &    3.992 $\pm$ 0.005 &    -0.0334 $\pm$ 0.0004 \\
       & $\Lambda$CDM & Freq. &    0.438 $\pm$ 0.003 &  0.1709 $\pm$ 0.0004 &  0.1182 $\pm$ 0.0011 &    4.217 $\pm$ 0.026 &  -0.03276 $\pm$ 0.00014 \\
       &              & Bay. &    0.440 $\pm$ 0.007 &  0.1706 $\pm$ 0.0015 &  0.1168 $\pm$ 0.0026 &  3.9956 $\pm$ 0.0024 &    -0.0323 $\pm$ 0.0008 \\
\bottomrule
\end{tabular}
\label{tab:bayfreq}
\end{table*}

\subsection{Profile Likelihood}

The (non-marginalised) profile likelihood at\footnote{Again, this value was chosen for comparison with \citet{Dam_2017}.} $z \Ns {CMB} = 0.033$ gives the maximum likelihood value for the cosmological parameter as $f\Ns{v0, MLE} = 0.675$, and $\Omega\Ns{M0, MLE} = 0.433$ (\cref{fig:profilelike}) when maximising the nuisance parameters, as seen by
\begin{equation}\label{eq:prof}
    \mathcal{L}\Ns{p}(\Theta) = \max_{\phi} \mathcal{L}(\Theta, \phi),
\end{equation}
where $\mathcal{L}$ is the likelihood as defined in \cref{eq:likelihoodFull} in terms of the nuisance parameters $\phi$, and $\Theta$ is constrained to take a range of values. The profile likelihoods have similarities to the JLA likelihoods from \citet[Fig.~3b]{Dam_2017} in shape, however, the relative sizes of the likelihoods differ between the JLA and \pplus\ surveys. 
The JLA likelihood seems to favour the timescape cosmology. while the \pplus\ likelihood, by contrast, favours the \lcdm\ cosmology. The integrated likelihood of the \lcdm\ is 1.32 times higher than timescape for the P+1690 sample with $z \Ns {min} = 0.033$ and 1.18 times as high for $z \Ns {min} = 0.075$. In particular, the likelihood ratio decreases when restricting to the \sne\ beyond the SHS$_\alpha$. 

The MLE value for the cosmological parameters is in the broad 1–2 $\sigma$ bounds chosen for the Bayesian analysis. \cref{fig:profilelike} shows the likelihood for the P+1690 sample and the random subsamples which are normalised for comparison. When we consider the entire sample, the likelihood exhibits a narrower Gaussian profile when contrasted with the subsamples of size $N=1000$.

\subsection{Statistical Approaches}\label{sec:bayfreq}

\begin{figure}
	\includegraphics[width=\columnwidth]{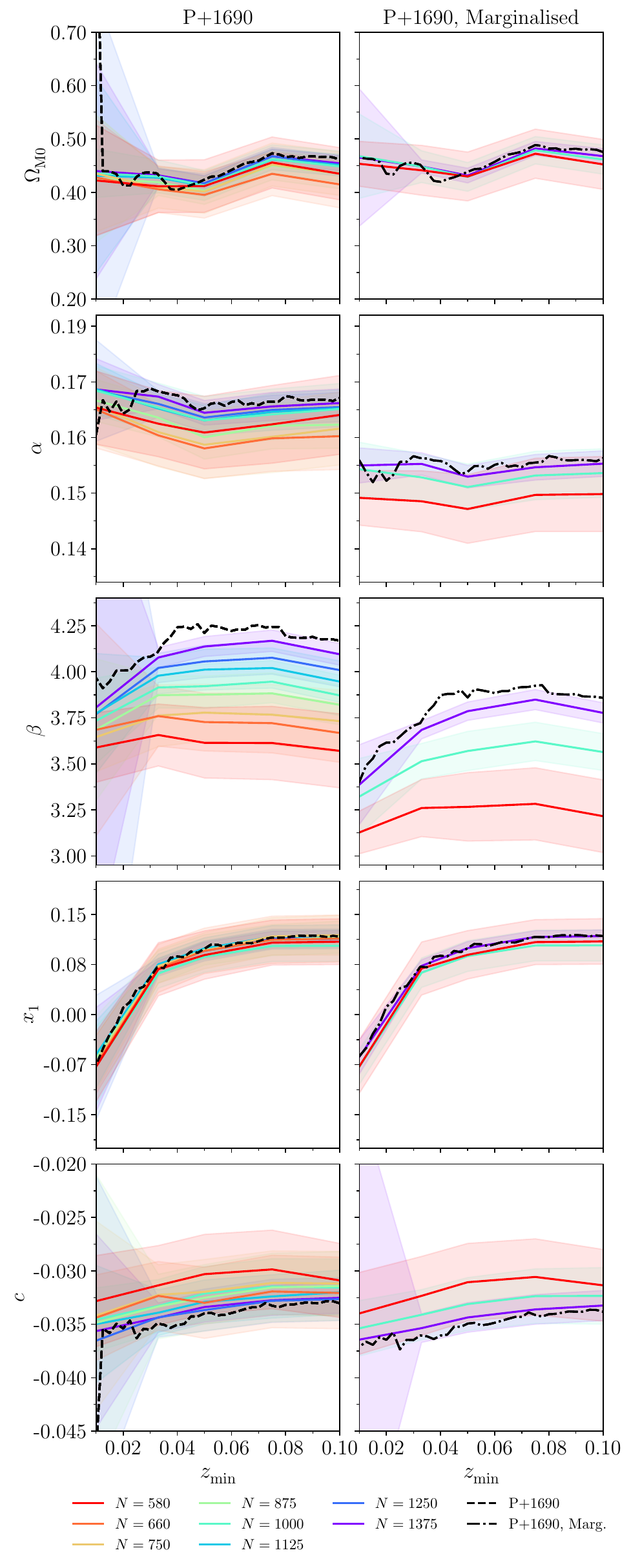}
    \caption{Median and standard deviation of 50 random subsamples of various sizes from the \pplus\ catalogue for the timescape model, based on the non-marginalised (\cref{eq:likelihoodFull}, left column) and the marginalised (\cref{eq:likelihoodMarg}, right column) likelihoods to show how the different parameters change with sample size. In the first row, we display the effective dressed parameter $\Omega_{\rm M0} = \frac12 (1 - f\Ns{v0}) (2 + f\Ns{v0})$ from \cref{eq:omegam_ts}.}
    \label{fig:modelparamsRS}
\end{figure}

In our analysis, we considered the frequentist approach using the marginalised (\cref{eq:likelihoodMarg}) and non-marginalised (\cref{eq:likelihoodFull}) likelihoods as well as the Bayesian evidence. Because of the computational resource requirements associated with marginalising the likelihood, we exclusively performed the latter based on \cref{eq:likelihoodFull}. A comparison of the Bayesian and frequentist approach based on the cosmological model parameters is shown in \cref{fig:omega}, while the differences between the marginalised and non-marginalised frequentist results are depicted in \cref{fig:modelparamsRS}. The values beyond the SHS$_\alpha$ for all parameters, models and statistical methods are given in \cref{tab:bayfreq}.

In the frequentist approach, the marginalised likelihood \cref{eq:likelihoodMarg} maintains the same shape as the non-marginalised treatment, with different numerical values (as shown in the right columns of \cref{fig:random,fig:modelparamsRS,fig:conv}). In particular, the results found for $\alpha$ and $\beta$ are significantly lower in the marginalised case. Furthermore, the marginalised likelihood effectively manages the dispersion within the low-redshift range (see \cref{sec:reducelowz}), except with low values of sample size ($N$).

For a quantitative analysis of the statistical approaches, we consider the median and standard deviation of the model and light-curve parameters beyond the SHS$_\alpha$ (\cref{tab:bayfreq}). For the light-curve parameter values $\alpha, x\Z 1, \beta$ and $c$, there are no discernible differences between the two statistical approaches nor for the two models.

Beyond the light-curve parameters, however, it is essential to note that the timescape model depends on the parameter $f\Ns{v0}$ according to \cref{eq:omegam_ts}, while \lcdm\ considers $\Omega\Ns{M0}$.
The marginalised results for these exhibit minimal deviation from the non-marginalised approach, as visually depicted in the top row of \cref{fig:modelparamsRS}. 
While the cosmological parameter values are consistent between each statistical approach and the samples, we find expected deviations between timescape and \lcdm\ for each respective sample. Timescape utilises an effective dressed parameter, calculated from $f\Ns{v0}$, and is not directly related to the $\Omega\Ns{M0}$ of the \lcdm\ model. In particular, because of various model limitations, we emphasise that these values are not used to constrain the \lcdm\ parameters.

Furthermore, these parameters consistently agree within 1–2$\sigma$ for the frequentist and Bayesian values across different values of $\zmin$ (\cref{fig:omega}).
This emphasises that these two statistical methods are consistent.
A more comprehensive statistical comparison of the two cosmological models is carried out in \cref{sec:modelcomp} in terms of a Bayesian analysis.

\subsection{Model Comparison}\label{sec:modelcomp}

\begin{figure}
	\includegraphics[width=\columnwidth]{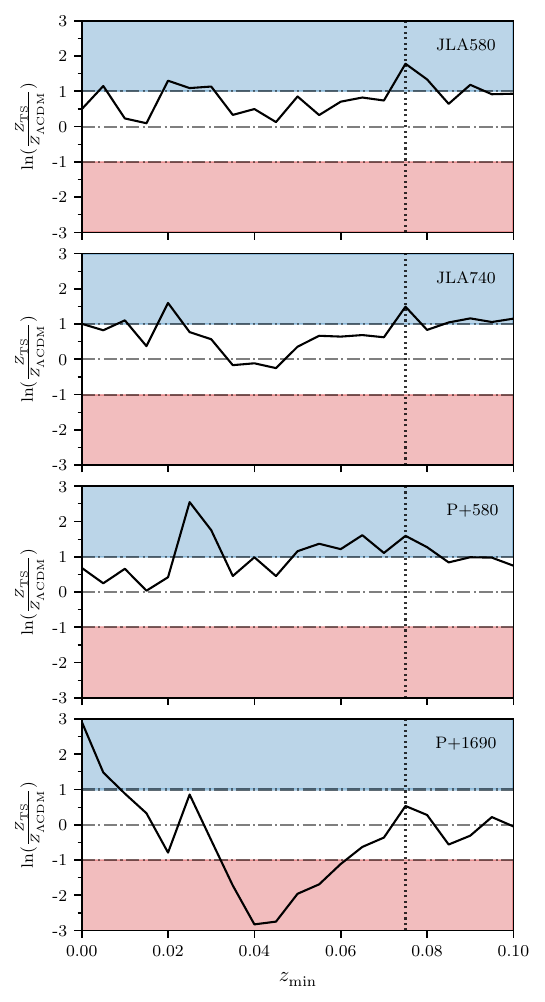}
    \caption{Bayesian evidence $\ln{B}$ of the timescape model relative to the \lcdm\ model when data $z<\zmin$ is progressively removed and parameters are refitted for the entire remaining sample $z\ge \zmin$. {Note that while the Bayesian analysis does yield error bars for the evidence, they are too small to be visible in this plot, as the error for each importance nested sampling log-evidence datapoint is $\sim0.013$.} 
    {The shaded regions show positive evidence on the Jeffrey's scale \citep[\cref{sec:statmethods} and][]{Kass_1995}, for the timescape (upper, blue) and \LCDM\ (lower, red) models.}
    The observed variation with $\zmin$ strongly suggests the presence of additional systematic effects as expected in generic inhomogeneous cosmologies, which will be presented in a forthcoming analysis. The vertical dotted line represents the SHS$_\alpha$ found at $z\Ns{CMB}=0.075$. 
    % The overall evidence is $0.80 \pm 0.04$ for JLA580, $0.72 \pm 0.04$ for JLA740 and $1.02 \pm 0.04$ for P+580, all favouring the timescape model. In contrast, the integrated evidence for the P+1690 yields $-0.46 \pm 0.04$, which favours the \lcdm\ model.
    }
    \label{fig:modelcomp}
\end{figure}

We now present the synthesis of all previous results in terms of a Bayesian model comparison, shown in \cref{fig:modelcomp}. A Bayes factor for each pair of models is determined by a joint refit of the parameters  $\{\alpha,x\Z1,\beta,c \}$ on the r.h.s.\ of \cref{eq:bandaid} for light-curves and the energy--density fractions $\{\Omega \Ns {M0}\}$ (\LCDM) or $\{\fvn\}$ (timescape) that feed into $\dL$ on the r.h.s.\ of \cref{eq:mu}. These correspond to empirical observations and cosmological models\footnote{The empty Milne universe has no free cosmological parameter, given that the absolute magnitude $M\Ns B$ and Hubble constant $\Hn$ are marginalised over as nuisance parameters. As discussed by \citet{Dam_2017} this complicates any Bayesian comparison. The Milne universe is of mathematical interest as a limiting case but is not physically viable and is omitted in this paper.} respectively.

Within each panel, parameters (\cref{eq:bandaid,eq:mu}) for all $z>\zmin$ \sne\ are refit as $z<\zmin$ events are discarded, with $\zmin$ treated as a running variable.

The P+1690 panel has significant differences from the other panels. The differences between the panels are nonetheless entirely consistent with the expectations of the timescape cosmology, once all statistical distribution biases are accounted for.

Since both \LCDM\ and timescape are compared here with precisely the same \sne\ data, the Bayes factor depends predominantly on how the theoretical models depart from average isotropy and homogeneity. As discussed in \cref{sec:cosmo} in \LCDM\ -- or indeed in any FLRW model -- one has globally defined {\em peculiar velocities}, whereas for viable inhomogeneous cosmologies, one must introduce the notion of non-kinematic differential expansion \citep{Bolejko_2016}. Thus, further work is required to include differential peculiar expansions in timescape to make fair Bayesian comparisons to \LCDM\ with peculiar velocities. To roughly quantify the magnitude of any changes, we repeated the analysis of \cref{fig:modelcomp}.

The CEP and quasilocal uniform Hubble expansion condition in the timescape cosmology focus on the distinguishability of motion from average expansion, and demand a reassessment of peculiar velocities in the timescape cosmology. Thus we deliberately excluded any peculiar velocity corrections as they would introduce a bias towards the \LCDM\ model. However, the omission of peculiar velocity corrections results in a reduction of the error bars in our analysis which has to be considered in the evaluation. Alongside the systematic covariance (c.f. \cref{sec:systcov}), careful attention must be given to calibration-related uncertainties beyond those considered in the covariance, as well as variations stemming from selection effects in the subsamples. Furthermore, it is expected that the \LCDM\ model is less accurate in the low-redshift regime, given its assumption of spatial homogeneity.

The specific feature of the timescape model which is compared here is the {\em uniform quasilocal Hubble expansion condition} which automatically dispenses with the Hubble tension \citep{Wiltshire_2007_clocks, Wiltshire_2008, Wiltshire_2009_obs}. It is the mathematical model for cosmic structure and emergence of statistical homogeneity which is the key to understanding the explanatory significance of \cref{fig:modelcomp}. With this insight in hand, a set of distribution effects are visible across the four panels of \cref{fig:modelcomp}, which we can group as follows.

\subsubsection{Sensitivity}\label{sec:sensitivity} The P+1690 Pantheon+ full sample (lowest panel) contains significantly more $\zlo$ {\sne\ }than in the upper 3 panels. Since
these additions to the P+1690 distribution are at {the} smallest cosmological distances where the variation of the \HL\ law is the greatest, we can expect to see a greater sensitivity of the Bayes factors in this region. The results {can} all {be} consistently {interpreted in light of} the quasilocal uniform Hubble expansion hypothesis as follows:

\begin{enumerate}[label=(\roman*), topsep=0pt, itemsep=0pt,leftmargin=*, align=right] 
    \item Including the closest $\zmin\lsim0.01$ \sne\ drives $\ln B$ evidence for the full sample from {the stage of {``not worth more than a bare mention''} on the Jeffrey's scale \citep[\cref{sec:statmethods} and][]{Kass_1995}} to positive evidence in favour of timescape, and just below the $\ln B\goesas3$ threshold at which the Bayesian evidence is strong by Jeffrey's scale \citep{Kass_1995}. It is the value of the $\zmin=0$ intercept, $\ln B=2.93$, which drives the vertical scale of \cref{fig:modelcomp}, {which aligns} with the expectation for a typical galaxy -- like the Milky Way -- in a thin filament in a void. 
    \item The peak in the P+580 panel ($\ln B = 2.55$) is significantly higher than in the JLA740 panel ($\ln B = 1.60$), both occurring within the range $0.02\lsim\zmin\lsim0.03$ or $60\hM\lsim \dL\lsim95\hM$. Since exactly the same \sne\ events are involved, this {interpretation} is consistent with:
    \begin{enumerate}[label=\alph*), left=1em] 
        \item the more stringent colour selection criteria in Pantheon+;
        \item the $\zmin=0.023$ cutoff used in the \cite{Riess_2007} Gold07 MLCS2k2 analysis and results of the first Bayesian \LCDM\ /  timescape comparison using that sample \citep{Leith_2008};
        \item convergence (see \cref{sec:convergence} below);
        \item the expected transition scale from the nearby nonlinear regime to the larger scale linear regime;
        \item the hypothesis that this feature is correlated with the Centaurus--Hydra complex, i.e., the {``Great Attractor''} \citep{Wiltshire_2013_rest, McKay_2016}.
    \end{enumerate}
    \item\label{num:lnB-pp1690} The minimum $\ln B$ value in the P+1690 panel at $\zmin = 0.04$ is significantly lower than in the P+580 panel, with a Bayes factor $\ln B = -2.82$.
    {This value indicates the P+1690 sample favours the \LCDM\ model {within} this regime. However, {when} comparing the value to the P+580 subsample, we find a Bayes factor of $\ln B = 0.98$ in favour of timescape at} 
    the same redshift cut. {This is likely} indicative of a distribution sampling bias {which is reaffirmed} by the ordering $\ln B|\Ns{P+580}>\ln B|\Ns{JLA580}>\ln B|\Ns{JLA740}>\ln B|\Ns{P+1690}$. {Within this interpretation, these results are consistent with:}
    \begin{enumerate}[label=\alph*), left=1em] 
        \item\label{num:lnB-colour} the more stringent colour selection criteria in Pantheon+;
        \item\label{num:lnB-stats} the hypothesis that the feature is a statistical distribution bias resulting from the degeneracy of $\beta c$ with $\dL$ when $M\Ns B$ and $\Hn$ are marginalised over in \cref{eq:tripp} as nuisance parameters;
        \item\label{num:lnB-conv} convergence (see \cref{sec:convergence} below);
        \item\label{num:lnB-bootes} the hypothesis that this feature is correlated with structures leading to a puzzling bulk flow in the \LCDM/FLRW framework 
        {situated along the $\{\ell,b\}=\{(297/117)\degree\pm4 \degree,(-6/6)\degree\pm3\degree\}$ axis \citep{Watkins_2023,Whitford_2023}. In the timescape framework, the different redshift regimes along this axis can be reinterpreted as local structures. In particular, the regime $z\goesas0.033$--$0.06$ can be interpreted as fine structure of the cosmic web and the foreground structures through which {they} are observed.} 
        Identifying structures along this axis is a challenge due to obscuration by our galaxy (Zone of Avoidance) compounding difficulties associated with the identification of voids.
        The Chamaeleon overdensity in the South Pole Wall centred at a redshift of $\sim0.04$ \citep{Pomarede_2020} is one overdensity matching what we {might} denote a {\em fine structure distance tension}. 
    \end{enumerate}
    \item\label{num:lnB-highz} 
    The $\ln B$ values for $\zmin>0.075$ to the right of the vertical dotted line are close to $\ln B=0$ for the full P+1690 sample, but are consistently at the marginally positive boundary $\ln B=1$ for all other cases. This is consistent with \cref{num:lnB-colour,num:lnB-stats,num:lnB-conv} of \cref{num:lnB-pp1690} above. 
\end{enumerate}

\subsubsection{Convergence}\label{sec:convergence} In \cref{fig:conv,fig:modelparamsRS} bootstrap methods were applied to test the convergence of the random subsamples of $N$ \sne\ of the full P+1690 Pantheon+ as $N\to1535$ \sne\ (the P+1690 sample consists of 1535 unique \sne). As was already discussed in \cref{sec:shs} this {could} be due to distribution biases arising from structures in the range $0.04\lsim\zmin\lsim0.06$. In \cref{fig:modelcomp} these effects {appear to be} associated with
\begin{enumerate}[label=(\roman*), topsep=0pt, itemsep=0pt,leftmargin=*, align=right] 
    \item the height of the $\ln B$ peak value in the JLA580 and P+580 panels;
    \item the magnitude of the minimum $\ln B$ in the P+1690 and P+580 panels;
    \item\label{num:conv-highz} $\ln B$ values for $\zmin>0.075$  (see \cref{sec:convergence}, \cref{num:lnB-highz} above).
We note that in the timescape cosmology it is expected that for very large\footnote{Here, {the term ``very large''} is determined by the radius of convergence of a Taylor series expansion of the distance--redshift law relative to the precision of datasets used.} $\zmin\gg0.033$, $\ln B\to0$. The fact that the upper 3 panels of \cref{fig:modelcomp} have $\ln B\to1$ is consistent with:
    \begin{enumerate}[label=\alph*), left=1em] 
        \item $\zmin\goesas0.1$ being a scale at which very small corrections to average expansion law from inhomogeneities can still be detected;
        \item the fact that Malmquist distribution biases have thus far not been subtracted in our analysis.
    \end{enumerate}
\end{enumerate}

It should be noted that while these conclusions are consistent with the results obtained, they could also agree with other interpretations. The key finding that our confirmation of convergence in bootstrap resampling is associated with the value of $\beta c$ in \cref{eq:tripp} is deferred to \cref{sec:discussion}.

\subsubsection{Novel Selection Criteria}\label{sec:novel_selection_criteria} \cref{sec:sensitivity,sec:convergence} have features that suggest further refinement of our empirical definition of the SHS$_\alpha$ statistic is likely to improve the subtraction of peculiar velocities/expansion in very large distance--redshift surveys. The clearest signal for this are the systematic differences of $\ln B$ values for $\zmin>0.075$ and the points made in \cref{sec:convergence}, \cref{num:lnB-pp1690} above.

In particular, now that an old debate \citep{Nielsen_2016, Rubin_2016, Dam_2017, Colin_2020} is resolved, we see that the most significant aspects of each viewpoint were correct within the context of assumptions about unknown systematics. Each researcher in the debate brought different perspectives -- whether instrumentational, phenomenological or theoretical. Systematics due to the interplay of model assumptions about cosmic expansion and calibration of the entire available \sne\ catalogue are key to understanding the convergence of $\alpha x\Z 1 \to 0.0204 \pm 0.0010$, which defines SHS$_\alpha$. Its convergence can be seen in \cref{fig:shs}, in contrast with previous debates that focussed on SHS$_{\rm g2}$, which fails to converge for the samples considered. In particular, the $\beta$ parameter from the light-curve analysis does not converge as $N\to N_{\rm u}=1535$ (for the full P+1690 catalogue) as seen in \cref{fig:conv} which leads to a lack of convergence for the SHS$_{{\rm g}2}$.

Inhomogeneity from cosmic structure underlies the key systematic interpreted as peculiar velocities in the \LCDM\ model and \textit{peculiar expansion} in the timescape model, and which drives the Hubble tension \citep{Di_Valentino_2021}.

Peculiar velocity modelling based on FLRW assumptions is routinely included in many \sne\ data releases. We needed to remove such modelling for our non-FLRW analysis of \pplus. While our Bayesian comparison and interpretation appears sufficient to resolve the largest contributions to the Hubble tension, \cref{fig:shs} shows that new selection criteria based on the SHS$_\alpha$ should already be sufficient to isolate smaller higher order effects of cosmic inhomogeneity in the \pplus\ catalogue.
This is supported by \cref{fig:modelcomp}: for P+1690 we find $\ln B \sim 0$ beyond the SHS$_\alpha$ instead of $\ln B \sim 1$ in P+580.

The most significant result is that the overall Bayesian evidence from bootstrapped resampling of \pplus\ takes a value $\ln B= 1.02\pm0.04$ in favour of timescape {in the P+580 sample}. {When considering the P+1690 sample, the integrated Bayesian evidence marginally favours \LCDM, however, the evidence significantly changes across $\zmin$, resulting in a redshift regime that positively favours timescape} on the Jeffrey's scale {and another with positive evidence for \LCDM\ (c.f. \cref{fig:modelcomp}). To evaluate the evidence of the full sample, selection effects from the redshift distribution and systematics} such as Malmquist bias have to be accounted for using the broadened definition of model independence. 

It should be stressed that this is completely independent of the goodness of fit of the timescape cosmology. 

\section{Discussion}\label{sec:discussion}

The updated analysis of the timescape and \lcdm\ models, utilising new data of over twice the size, has uncovered significant discoveries regarding both the structure of the catalogue and the evaluation of the models.

The \pplus\ results use a covariance matrix of distance moduli, assuming a fiducial $w$CDM cosmology. {To} compare the fundamental differences between cosmological models, or to identify any inherent biases in the current standard of cosmology, it is crucial to perform cosmology-dependent steps as late as possible in the data-reduction pipeline. Due to the cosmology-dependent bias-correction simulations that were introduced by \pplus\ and applied to the SALT2mu algorithm, along with the inclusion of peculiar velocity modelling in their final Hubble diagram, it was not possible to incorporate these calibrations in the calculation of the covariance matrix. Consequently, there remains the possibility of uncorrected selection biases in these results. However, we presume that these biases are of a similar magnitude for both models. This approach allows a fair comparison between cosmological models not built upon the FLRW geometry. {To} mitigate potential issues and biases, we cross-reference our results with the observations reported in the JLA catalogue by \citet{Dam_2017}.

{Nevertheless, it is important to consider potential additional systematics that we were unable to address in the covariance, along with selection effects from the subsamples. These factors should be taken into account during the evaluation and warrant further constraints to enhance the robustness of the analysis.}

Prior to the marginalisation process, we recover the differences in the SALT2 surfaces reported by \citet{Taylor_2021}, when comparing the JLA and \pplus\ catalogues. It is important to account for the inherent variations within the published datasets when making comparisons between different \sne\ catalogues.

As there are significant differences in the results of the \common\ and the full \pplus\ sample, we conduct the analysis for random subsamples (see \cref{sec:random} for more details) for both the JLA and the \pplus\ datasets to investigate the selection effects. The dispersion of the randomised results for the \pplus\ sample is greater than JLA, having significant scatter at low-redshift cuts. Notably, $42\%$ of unique \sne\ in the \pplus\ catalogue have a redshift of $z\Ns{CMB} \leq  0.075$. Given the large number of low-redshift observations, there is considerable scatter in the $x\Z 1$ and $c$ parameters for $\zmin <  0.0225$.
Hence, light-curve parameters deduced using these values must consider the uncertainties stemming from the variations in the observed parameters at low-redshift ranges. 

One reason for the scatter is the fact that the supernova observations within the \pplus\ catalogue are derived from diverse samples and thus the value adopted for $\Hn$ differs. To account for this, we include marginalisation over the zero-point offset in the definition of the likelihood function. Such an approach effectively controls the level of scatter.

As the redshift cut increases, the scatter becomes negligible. However, the light-curve parameters only converge beyond the scale of statistical homogeneity. To evaluate this quantitatively, we rely on a power--law method fitted to $\alpha x\Z 1$ presented in \cref{sec:shs} and quantify its approach to convergence.

For this data driven approach we take a percentage range of the $\alpha x\Z 1$ limiting value for high $\zmin$ which is motivated by a few key reasons. One rationale is the need for this method to be applicable to future datasets, and we cannot assume uniformity in the level of noise across diverse datasets. Our method is also more robust to datapoints that have not yet reached the SHS$_\alpha$ or display numerical instability.

Different levels of statistical noise may lead to earlier or later convergence. We chose to use 5\% as it provides more constraints on the SHS$_\alpha$ for the convergence while having a higher tolerance for noise and fitting constraints than a 3\% uncertainty does. In \pplus, the 5\% bound for the chosen range corresponds to 2.6$\sigma$ for $\alpha x\Z 1$ using a \lcdm\ model. However, because of the larger values of $\beta$ a $1$--$3$\% bound may be more appropriate for SHS$_\beta$ which was not possible to define here due to the lack of convergence.

Nonetheless, this approach does exhibit several limitations. Firstly, power--law functions tend to approach negative infinity as $x\Z 1 \to 0$ and it is obvious that the $x\Z1$ parameter remains finite and bounded across all redshifts. This highlights the important need for a well-defined lower bound when applying this model-fitting method. Modifying the power--law by introducing a fourth parameter, such as $-a(z-z\Z 0)^{-n} + C$, does marginally improve the $\chi^2$ statistic and, for negative values of $z\Z 0$ prevents the power--law from diverging for $x\Z 1 \to 0$. However, this modification does not significantly alter the resulting SHS$_\alpha$ and instead increases the Akaike information criterion (AIC)\footnote{The Akaike information criterion (AIC) is a criterion for the evaluation of statistical models based on the goodness of fit and the number of estimated parameters. For more information see \citet{Akaike_1974} and \citet[Appendix~C]{Dam_2017}.}.

The second limitation pertains to the fitting process itself. While fitting a power--law, any numerical spikes present in the data can significantly influence the resulting best-fit curve.

Lastly, the power--law method relies on the convergence of the $x\Z 1$ parameter, which was not the case for JLA. The $x\Z 1$ convergence issue introduces a challenge when comparing results across different catalogues that may not control the redshift dependence of $x\Z 1$ as effectively as \citet{Brout_2022_cal} did in \pplus. Due to the degeneracy of $\alpha$ and $x\Z 1$, convergence issues within one of the parameters can be counterbalanced by the other, making the treatment more robust when considering the product.

As discussed previously in \cref{sec:shs} and shown in \cref{fig:ext_z_c}, the parameters $\beta$ and $c$ do not converge and thus a SHS$_\beta$ based on these is not considered as an alternative to the SHS$_\alpha$. This is likely due to the Malmquist biases present at redshifts greater than $\zmin \sim  0.05$ which we could not yet account for in the timescape model. Therefore, we also exclude the correction from \lcdm\ for consistency. Applying corrections for this bias to timescape would be an important step to improve our sample and confirm the SHS$_\alpha$ by a second set of parameters. 

The result we obtained for the SHS$_\alpha$ with the power--law method significantly differs from the one found by \citet{Dam_2017}. These differences stem from inherent biases within the distinct datasets, emphasising the need for robust controls to assess their convergence behaviours effectively. Such calibration was done effectively by \pplus\ for the $x\Z 1$ parameter. Further cosmology-independent treatment of such issues is needed for a more comprehensive examination of the SHS$_\alpha$.

Moreover, random sampling has a tangible impact on the parameter values. For $\alpha$, $x\Z 1$ and $c$, these changes are not significant and the results converge towards that of the full sample when increasing the sample size. However, the most significant change is in the $\beta$ parameter for varying subsample sizes. For this parameter, we do not observe a convergence, as the final value differs significantly from the ones found for lower $N$. Considering marginalised likelihoods does not resolve this feature as it changes the absolute values but not the behaviour for varying sample sizes. Consequently, the results for $\beta$ cannot be attributed to the choice of $\Hn$. As shown in \cref{fig:conv}, selecting subsamples with different redshift distributions resolves this issue. Therefore, for a thorough investigation into the convergence of the light-curve parameters, a sample with uniform redshift distribution would be needed. Constructing such a sample will require more \sne\ particularly at {$z\Ns{CMB}\gsim0.04$}.

A noteworthy observation regarding the light-curve parameters of the \pplus\ dataset, specifically $x\Z 1$, is that they converge for $\zmin = 0.075^{+0.007}_{-0.009}$. In contrast, the dependency on redshift for the light-curve parameters in the JLA catalogue indicates the presence of an accelerated expansion. This feature has been in heavy debate for the JLA sample. \citet{Nielsen_2016} enforced redshift independence in $x\Z 1$ and $c$ in their analysis, which was criticised by \citet{Rubin_2016}. However, \citet{Rubin_2016} added twelve additional parameters to the Bayesian Hierarchical Model, and their inclusion has been challenged by several papers, notably, \citet{Dam_2017} and \citet{Colin_2020}. 
\citet{Dam_2017} found that the {introduction of additional parameters} did not increase $\ln B$, while \citet{Colin_2020} {criticised the addition made by \citet{Rubin_2016} from the principles of} the Bayesian information criterion (BIC)\footnote{The Bayesian information criterion (BIC) is closely related to the Akaike information criterion (AIC) but introduces a larger penalty term for $k \geq 8$. The BIC is given by: ${\rm BIC} = k\ln{N} - 2\ln{\mathcal{(L)}}$. For more information see \citet{Kass_1995} and \citet[Appendix~C]{Dam_2017}.}. 
When the same supernova analysis as in \citet{Dam_2017} is conducted on the \pplus\ data, the lack of redshift dependence indicates that these features from JLA were due to different calibration systematics and incomplete datasets. In particular, since $x\Z 1\to\w{const.}$ for $\zmin > \w{SHS}_\alpha$, the systematic error in the JLA catalogue that was at the centre of the debate has been controlled in \pplus.

The Bayesian analysis of both the JLA data and the \common\ agree with the results by \citet{Dam_2017}. When we extend the analysis to encompass the complete P+1690 dataset, certain artefacts emerge, stemming from distribution-related effects. In particular, the \pplus\ catalogue incorporates a substantial number of low-redshift observations that fall below the SHS$_\alpha$. In this low-redshift regime, the fundamental assumption of statistical homogeneity does not hold, and consequently, the timescape model demonstrates a more favourable fit to the data.
The P+1690 global minimum $\ln B\Ns{P+1690}=-2.82$, (c.f., \cref{sec:convergence}, \cref{num:conv-highz}) {can be interpreted either in favour of the \LCDM\ model or as a distribution effect from the redshift distribution of the observations, as the other analysed samples favour the timescape model.}
This preference arises due to specific structural features present in the data and the partial reliance on the \lcdm\ cosmological framework to account for these features. 

Future \sne\ surveys such as those from Euclid, the Rubin Observatory, and the Roman Space Telescope will vastly expand on the \sne\ sample size and allow for more detailed examination of large scale structures \citep{Euclid_2020_forecast,Euclid_2022_wide,Rubin_Observatory_2019,Roman_Telescope_2021}. For these detailed analyses to be possible, the data products from the surveys must be presented free from underlying cosmological models, as was achieved here with \pplus. For example, subtle biases that favour \lcdm\ can be introduced when correcting for sample related biases such as the Malmquist bias. 

\section{Conclusions} \label{sec:conclusions}

We have performed a from--first--principles reanalysis of the best available \sne\ catalogue, Pantheon+ \citep{Scolnic_2022}. In view of our results, the \pplus\ data may already contain a sufficient wealth of data to support a fundamental change in thinking about the \LCDM\ model as the standard model of cosmology. Indeed, the possibility of such a paradigm shift is supported by new results released by both the Dark Energy Spectroscopic Instrument (DESI) \citep[DESI,][]{desi_2024} and the Dark Energy Survey \citep[DES,][]{des_2024}, as well as our own further analysis \citep{Seifert_2024}.

Supernovae Ia were the events that first convinced a majority of the community of the need for an observational paradigm shift \citep{Riess_1998, Perlmutter_1999}. That paradigm shift invoked a mysterious {\textit{dark energy}}, which has always been recognised as a place-holder for new physics. All major paradigm shifts in physics have involved the complex interplay of concepts, experimental phenomenology and mathematical logic. While such questions are likely to still play out for decades, {in our view some new physics features of the dark energy place-holder may already be} emerging.

The observed spatial isotropy and homogeneity of the most distant sources has been a deep theoretical conundrum ever since relativistic cosmology was first constructed a century ago. By necessity, the first mathematical cosmologists put the answer into their models as an assumption, and elevated the assumption to the Cosmological Principle. Rather than assuming spatial isotropy and homogeneity, any appeal to a deeper physical principle should seek to explain such puzzling observations. The timescape cosmology {seeks to do} this by returning to fundamental questions left unanswered in the first formulations of relativistic cosmology, via its Cosmological Equivalence Principle (CEP) \citep{Wiltshire_2008}, an extension of the suite of Einstein's equivalence principles. {In contrast to modified gravity models whose new physics directly violates either the Weak or Strong} Equivalence Principles, in the timescape model each of Einstein's principles remains intact and embedded within the others according to its scale of validity.

{For our purely empirical approach, the phenomenology of the quasilocal uniform Hubble expansion condition provides a means} for operationally determining the relevant scales SHS$_\alpha$, SHS$_\beta$ -- as required by the new selection criteria that emerged from our data-driven bootstrapping in \cref{sec:convergence,sec:novel_selection_criteria}. As a direct consequence of our analysis, timescape's proposed resolution of the Hubble tension\footnote{A potential resolution of the Hubble tension is a generic mathematical feature of inhomomogeneous GR cosmologies \citep{Bolejko_2018}. The important question for the timescape scenario has been: Is it compelling?} has uncovered deeper understanding of many issues that may signal the beginning of a paradigm shift.

In the conceptual framework of \citet{Kuhn_1962}, paradigm shifts occur over timescales of decades to centuries when multiple failures of incremental research are followed by the resolution of anomalies through the novel predictions of a new paradigm. Since the preconditions of a genuine paradigm shift in cosmology are already well documented \citep{Aluri_2022, Peebles_2022}, let us propose the \begin{center}
\begin{minipage}{0.8\columnwidth}
    {\em null hypothesis} that:
\end{minipage}\vspace{2pt}
\begin{minipage}{0.7\columnwidth}
    \raggedright the {\em timescape resolution of the Hubble tension} (TRHT) is {\em not} accompanied by any novel selection criteria in fundamental cosmology.
\end{minipage}
\end{center} 
Following \citeauthor{Popper_1935}'s (\citeyear{Popper_1935}) criterion, 
we should seek just one compelling falsification of the null hypothesis to rule it out. We may verify the TRHT in this manner, but not ever prove it. 

As yet, we are not able to rule out the timescape. Rather the following quantitative observations of the TRHT resolve puzzles noted over the past 25 years with smaller sparser datasets:
\begin{itemize}[topsep=0pt, itemsep=0pt,leftmargin=*, align=right]
    \item  The results of \citet{Smale_2011_sne} showed slight Bayesian preference for timescape over \LCDM\ using MLCS2k2 and slight Bayesian preference for \LCDM\ over timescape using SALT/SALT2 fitters as then understood. The maximum evidence in {favour} of either model had $|\ln B|\lsim1.68$, depending on precise $\zmin$ cuts, which outliers were excluded etc. Detailed discussion of two decades of advances in light curve fitting, and the historical treatment of systematic uncertainties, is beyond the scope of this paper. Nonetheless, looking into details one finds, e.g., that with further analysis a sequence of orderings of $|\ln B|$ analogous to \cref{sec:sensitivity} \ref{num:lnB-pp1690} is to be found in \citet[Tables 6--9, Fig.~8]{Smale_2011_sne}.
    \item \citet{Smale_2011_sne} also discussed the fact that for timescape to be preferred in Bayesian analysis, the extinction by dust in the Milky Way is needed to be typical compared to other galaxies, which is in agreement with independent direct measurement of \citet{Finkelman_2008, Finkelman_2010}. In two studies of nearby galaxies these authors found values of the reddening parameter $R\Ns{V}=2.82 \pm 0.38$, $R\Ns{V}=2.71 \pm 0.43$ for their two samples of 7 and 8 galaxies respectively, indicating that the Milky Way value $R\Ns{V}=3.1$ was within a standard deviation of expectation. The Milky Way value had been used in the MLCS based values, but was not included as a parameter in the SALT/SALT2 methods leading to potential degeneracies with free parameters. 
    
    While direct comparison of the numerical value for the $\beta$ parameter that was favoured over a decade ago is complicated by the advances in light curve fitting, 
    what remains consistent is the fact that higher relative values of the $\beta$ parameter were favoured by timescape as compared to \LCDM\ \citep{Smale_2011_sne}. In the present analysis, the resulting $\beta$ values are much higher than considered previously. However, the \LCDM/timescape differences are negligible compared to the deviations caused by choosing different statistical approaches (\cref{tab:bayfreq}). In particular, using a likelihood that marginalises over the zero-point offset degenerate with the Hubble constant, significantly reduces the value of $\beta$. This is thus related to the Hubble tension. 
    \item Furthermore, for the timescape model to be a consistent description of the sample, an apparent ``Hubble bubble'' -- as mooted by \citet{Zehavi_1998} -- should be present in the data, as discussed by \citet{Smale_2011_sne}. Conceptually this strikes at the heart of the peculiar motion versus expansion dichotomy that the timescape cosmology resolves. {This is likely} related to the $\beta$ values obtained by the different models and statistical analyses, which in turn may resolve the Hubble tension, {as previously discussed.} 
\end{itemize}\vspace{2pt}
Our in depth comparison between the \lcdm\ and timescape cosmological models was only possible with \pplus\ as we were able to remove all cosmology-dependent calibration and correction steps. Without the standard corrections, biases such as the Malmquist bias are present in our analysis, however, we expect the resulting differences to be small when comparing models. From this extensive dataset of 1690 \sne\ observations we follow the analysis of \citet{Dam_2017} and construct a covariance matrix from the SALT2 parameters augmented with relevant perturbations and the statistical covariance from \pplus.

In our analysis for the parameters $\alpha$, $x\Z 1$ and $c$, we find significant differences between JLA and \pplus. We find a convergent behaviour of $x\Z 1$ with $\zmin$ due to resolved systematic errors which resolves a previous debate \citep{Nielsen_2016,Rubin_2016,Dam_2017,Colin_2020}. In contrast, the $\beta$ parameter does not converge for varying sample size.
To fully understand the nature of the $\beta$ convergence further tests are required with \sne\ datasets that have a uniform redshift distribution.

Furthermore, we consider the SHS in terms of a data-driven definition of SHS$_\alpha=0.075^{+0.007}_{-0.009}$ which is substantially larger than the conventional SHS$_{{\rm g}2}\goesas0.033$. By relying solely on observations, our approach gives a {new} representation of the underlying structure in the Universe. In particular, the conventional 2-point galaxy correlation function measure SHS${{\rm g}2}$ {identifies broad characteristics} of a transition towards average homogeneity, which coincides with {recognised} large-scale structures like the Great Attractor. However, the {SHS$_\alpha$} statistic {could expose additional} {finer structures} within the transition to homogeneity at scales corresponding to more distant structures.

For future analyses of the SHS, an additional consideration of the SHS$_\beta$ would enable a more thorough investigation. For this to be possible, a treatment of the Malmquist bias, which was unaccounted here, is needed.

Considering the Bayesian comparison of the two cosmological models, we find that timescape is positively favoured in the JLA740, JLA580 and P+580 datasets. However, the extension of the full \pplus\ catalogue reveals important artefacts that {suggest we further analyse} the underlying cosmic structures through which \sne\ {are observed.} The {evidence supporting} the timescape model at lower redshifts highlights the necessity of accounting for inhomogeneities{, even within an FLRW framework.}

The significant changes in the value for the $\ln B$ can be explained from the cosmic structures. To further investigate this and to determine which model is to be definitively preferred, more \sne\ at high redshifts are required to be calibrated in a cosmology-independent manner.

As final remarks we observe that one of the hallmarks of a genuine paradigm shift is that at tipping points, many puzzles tend to fall into place in rapid succession. This article is the first of several anticipated in the coming year. Some of the research directly related to the new characterisation of statistical homogeneity which is underway includes:
\begin{itemize}[topsep=0pt, itemsep=0pt,leftmargin=*, align=right]
    \item A deeper phenomenological understanding of the Newtonian limit in cosmology without a background FLRW geometry promises to be found. This may explain why, despite violation of the Buchert-Ehlers theorem \citep{Buchert_1997}, the heuristic computational scheme of \citet{Racz_2017} nonetheless produces results which are close to those of the timescape cosmology.
    \item Szekeres models \citep{Szekeres_1975} have a rich structure that can be used as effective models on $\goesas$SHS scales which can be directly constrained by large scale peculiar velocity / peculiar expansion datasets \citep{Hills_2022,OKeeffe_2022}. Future precision in such work relies on the definition of SHS$_\beta$ which was not yet possible due to the lack of cosmology-independent treatment of Malmquist bias. This is left for future work.
    \item  The BAO scale has been extracted directly from BOSS data (without FLRW assumptions) for the timescape cosmology \citep{Heinesen_2019}. This is important for future tests at scales of interest. However, calibration of this scale relative to the BAO at the CMB last-scattering surface requires understanding the interplay of general relativistic nonlinearities with matter at the galactic level. First steps to achieve this require understanding why particular simple classes of previously studied models and novel generalisations of those solutions fail \citep{galoppo2023topological}, leaving open the possibilities that may work in future \citep{Galoppo_2024a,Galoppo_2024b}.
\end{itemize}
\vspace{2pt}

Naturally, for the timescape model or any other non-FLRW model to be fully accepted by the community as an alternative to \LCDM, combined Bayesian evidence from independent tests is vital. The projections made by \citet[Fig.~10]{Sapone_2014} for performing the Clarkson-Bassett-Lu test \citep{Clarkson_2008} on Euclid mission data, require 1000 \sne\ in addition to Euclid BAOs to definitively distinguish the timescape expansion history from any FLRW expansion history. The goal of the present article was to answer the question for the \sne\ side of the Euclid projections.

Further development of BAO analysis is an urgent goal for fully testing the timescape model with Euclid. While direct extraction of the BAO scale from galaxy clustering data has been implemented for timescape \citep{Heinesen_2019}, calibration of the BAO requires more than a simple geometric analysis of the acoustic peaks in the Planck CMB data \citep{Duley_2013}. The angular scale of the acoustic peaks in Planck data is consistent with the expansion history that we have tested here with \sne. However, the ratio of baryons to nonbaryonic dark matter must be recalibrated\footnote{\citet{Camilleri_2024} report that Bayesian evidence in favour of timescape from Dark Energy Survey supernovae is reversed to very strongly favour \LCDM\ if BAO are included. However, their analysis makes purely geometric adjustments to the standard FLRW pipeline. Since incorporating detailed BAO analysis into the timescape cosmology entails revisiting matter model calibrations in the early universe, this result is encouraging for the timescape scenario as it suggests it may be falsifiable if its methodology is not fully implemented. For further discussion, see \citet{Seifert_2024}.} in the timescape model. Rather than simply changing the speed of sound in the primordial plasma, some fundamental questions relating to the statistics of quasilocal gravitational energy and angular momentum for bound systems are involved. In particular, whether or not cold dark matter particles are abundant, to what extent does a quasilocal collisionless pressure term act like an effective fluid made of particles? While the answer may still be years away, numerical simulations of the full Einstein equations \citep{Williams_2024}, combined with the plethora of new cosmological observations, may provide the best avenue for probing these fundamental questions.

General relativity is a deep and beautiful theory, whose principles remain to be fully explored -- and maybe extended -- to better understand the Universe. The hypotheses tested and further advanced here, if correct, indicate that we will likely be exploring those consequences for decades to come.

\section*{Acknowledgements}

We are deeply grateful to the Pantheon+ team for supplying the systematic perturbation files used in their analysis and for providing extensive feedback on its implementation. In particular, we thank Dillon Brout for his perceptive insights and hands--on assistance with our revisit of the Pantheon+ data reduction pipeline beyond conventional perturbed FLRW models. We also thank an anonymous referee for suggestions which greatly improved the paper. DLW, RRH and ZGL are supported by the Marsden Fund administered by the Royal Society of New Zealand, Te Apārangi under grants M1271 and M1255. RRH is also supported by the Rutherford Foundation Postdoctoral Fellowship under the grant RFT-UOC2203-PD. AS thanks the German Academic Exchange Service for supporting her stay at the University of Canterbury. DLW thanks Subir Sarkar for hosting a Visiting Oxford Erskine Fellowship, and Linacre College and the Rudolf Peierls Centre for Theoretical Physics for their generous support. We are indebted to Jenny Wagner for providing valuable assistance at the project outset and joining on in 2023, even though she did not have time to see it to its conclusion. We are grateful to all members of the University of Canterbury Gravity and Cosmology and Astrophysics groups 
for stimulating discussions: in particular, Quin Aicken Davies, Ethan Bull, Marco Galoppo, Christopher Harvey-Hawes, Morag Hills, Heather Sinclair-Wentworth, Shreyas Tiruvaskar and Michael Williams. We thank the University of Queensland group for valuable discussions, particularly Rianna Bell, Ryan Camilleri, Tamara Davis, Leonardo Giani, Rossana Ruggeri and Khaled Said. Finally, we thank Lawrence Dam, Asta Heinesen, Roya Mohayee, Eoin \'O Colg\'ain, Mohamed Rameez, David Rubin, Subir Sarkar, Dan Scolnic, Nathan Secrest, and Shahin Sheikh-Jabbari
for correspondence, discussions, debates and the dispassionate scientific consensus--building that made this analysis possible.

%%%%%%%%%%%%%%%%%%%%%%%%%%%%%%%%%%%%%%%%%%%%%%%%%
\section*{Data Availability}

A complete set of the codes and details used for our analysis and how to use them can
be found at
\citet{Seifert_2023}, and the covariance and input files are made available at \citet{Lane_2024}.

The JLA catalogue and their covariance can be found at \url{http://cdsarc.u-strasbg.fr/viz-bin/qcat?J/A+A/568/A22} and \url{https://supernovae.in2p3.fr/sdss_snls_jla/ReadMe.html} respectively.
The \pplus\ calibration/systematic perturbation files and data that were used in this paper were obtained via private communication with Dillon Brout and Dan Scolnic.

%%%%%%%%%%%%%%%%%%%% REFERENCES %%%%%%%%%%%%%%%%%%

\bibliographystyle{mnras}
\bibliography{references}

%%%%%%%%%%%%%%%%% APPENDICES %%%%%%%%%%%%%%%%%%%%%

\appendix

\section{Luminosity Distances}\label{app:model}
For the evaluation of the \sne\ observations we consider the distance moduli determined from the theoretical luminosity distances for the \lcdm\ and timescape models, respectively, following \cite{Dam_2017}. In both cases, the luminosity distances $d\Z L$ can be calculated from the model parameters and the distance moduli are given in terms of a Taylor expansion. For the \lcdm\ model, we have
\begin{align}
\dL&=\frac{(1+z)\CC}{\Hn\sqrt{|\Omega \Ns {k0}|}}\w{sinn}\left(\sqrt{|\Omega  \Ns 
 {k0}|}\int\limits^1
_{1/(1+z)}\frac{\dd y} {\mathcal{H}(y)}\right),
\nonumber\\ &\mathcal{H}(y)\equiv\sqrt{\Omega \Ns {R0}+ \Omega \Ns {M0}y+ \Omega \Ns {k0} y^2+\Omega \Ns {\Lambda 0} y^4}\,,
\nonumber\\ &
\w{sinn}(x)\equiv
\begin{cases} \sinh(x), &\Omega \Ns {k0}>0\\ x,&\Omega \Ns {k0}=0\\ \sin(x),&\Omega \Ns {k0}<0\\
\end{cases},\label{eq:dLlcdm}
\end{align}
with the present epoch values $\Omega \Ns {R0}$, $\Omega \Ns {M0}$ and $\Omega \Ns {\Lambda 0}$
of the radiation, non-relativistic matter and cosmological constant density
parameters, respectively. The condition to be spatially flat together with the Friedmann equation sum rule
\begin{align}
    \Omega \Ns {R} + \Omega \Ns {M} + \Omega \Ns {k} + \Omega \Ns {\Lambda} = 1
\end{align}
for all times and negligible $\Omega \Ns {R0}=4.15\times10^{-5}h^{-2}$ enables us to describe the model by two effective
free parameters, $\Hn$ and $\Omega \Ns {M0}\simeq1-\OmLn$.

In the timescape model \citep{Wiltshire_2007_clocks,Wiltshire_2007_sol,Wiltshire_2009_obs,Duley_2013}, the distance modulus is calculated differently and the observables considered are determined with respect to the finite infinity regions (\cref{sec:cosmo}). As described in \cite{Dam_2017}, the ``dressed'' luminosity distance $\dL$ is given by
\begin{align}
\dL &= (1+z)^2\dA,\label{dATS}
\end{align}
and the angular diameter $\dA$ can be determined from 
\begin{align}
\dA &=\CC\,t^{2/3}\int_t^{t\X0}
{\frac{2\,\dd \tb}{(2+\fv(\tb))(\tb)^{2/3}}}\nonumber
\\ &= \CC\,{t^{2/3}(\FF(t\Z0)-\FF(t))},
\label{dLTS}\\
\FF(t)&\equiv2t^{1/3}+{\frac{b^{1/3}}{6}}\ln\left(\frac{(t^{1/3}+b^{1/3})^2}{
t^{2/3}-b^{1/3}t^{1/3}+b^{2/3}}\right)\nonumber\\
&\hblank{20}+{\frac{b^{1/3}}{\sqrt{3}}}\tan^{-1}\left(\frac{2t^{1/3}-b^{1/3}}{
\sqrt{3}\,b^{1/3}}\right).\label{FF}
\end{align}
\begin{figure}
	\includegraphics[width=\columnwidth]{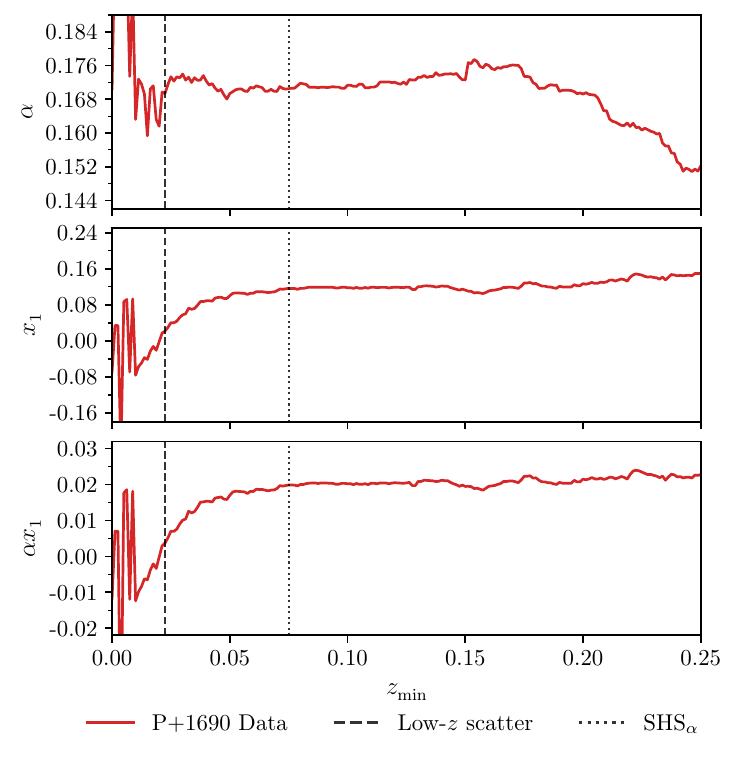}
    \caption{Similar to \cref{fig:ext_z_c}, we consider cuts at higher redshifts to analyse the evolution of the light-curve parameters and their degenerate products for the \lcdm\ model. As shown in \cref{fig:shs}, the product $\alpha x\Z 1$ can be fitted well with a power--law model and is thus used to determine the SHS$_\alpha$ (\cref{sec:shs}). In contrast to $\beta c$ (\cref{fig:ext_z_c}), it is not affected by Malmquist bias.}
    \label{fig:ext_z_x1}
\end{figure}
The volume-average time parameter, $t$, is defined implicitly in terms of
\beq
z+1=\frac{(2+\fv)\fv^{1/3}}{3\fvn^{1/3}\Hb t}
=\frac{2^{4/3}t^{1/3}(t+b)}{\fvn^{1/3}\Hb t(2t+3b)^{4/3}}\,,\label{redshift}
\eeq
where $b\equiv2(1-\fvn)(2+\fvn)/(9\fvn\Hb)$. The void volume fraction with the present epoch value $\fvn$ is given by
\beq\fv(t)=\frac{3\fvn\Hb t}{3\fvn\Hb t+(1-\fvn)(2+\fvn)}\,,
\eeq
and $\Hb$ is the ``bare Hubble
constant'' which is related to the observed
Hubble constant by $\Hb=2(2+\fvn)\Hn/(4{\fvn}^2+\fvn+4)$.

\newpage
\noindent The distance moduli are then given by\\
\begin{strip}
\begin{align}
&\mu\Ns{TS}=\mu\Z0(z)+\frac{5}{\ln10}\left\{
\left[\frac {24\,\fvn^{4}-23\,\fvn^{3}+99\,\fvn^{
2}+8}{2\left( 4\,\fvn^{2}+\fvn+4 \right) ^{2}}\right]z\right.\nonumber\\
&\hbox to 30pt{\hfil}-\left.\left[ 
{\frac {1984\,\fvn^{8}-4352\,\fvn^{7}+16515\fvn^{6}+14770\,\fvn^{5}+7819\,\fvn^{4}-11328\fvn^{3}+32080\,\fvn^{2}-128\,\fvn+960}{24\left( 4\fvn^{2}+\fvn+4 \right) ^{4}}}\right]z^2
+\dots\right\},\label{muTS}\\
&\mulcdm=\mu\Z0(z)+\frn5{\ln10}\left\{(1-\frn34\Omega \Ns {M0})z
-\left[\frn12+\frn12\Omega \Ns {M0}-\frn{27}{32}\Omega \Ns {M0}^2\right]z^2
+\left[\frn13-\frn18\Omega \Ns {M0}+\frn{21}{16}\Omega \Ns {M0}^2-\frn{45}{32}\Omega \Ns {M0}^3\right]z^3
+\dots\right\}.\label{mulcdm}%\\
\end{align}
\end{strip}

\noindent The term $\mu\Z0(z)\equiv25+5\log\Ns{10}[\CC z/(\Hn\w{Mpc})]=25+5\log\Ns{10}(2997.9\,h^{-1})+5\log\Ns{10}z$ which involves the Hubble constant $\Hn=100\,h\kmsMpc$ is common to all models.

\begin{figure*}
	\includegraphics[width=\linewidth]{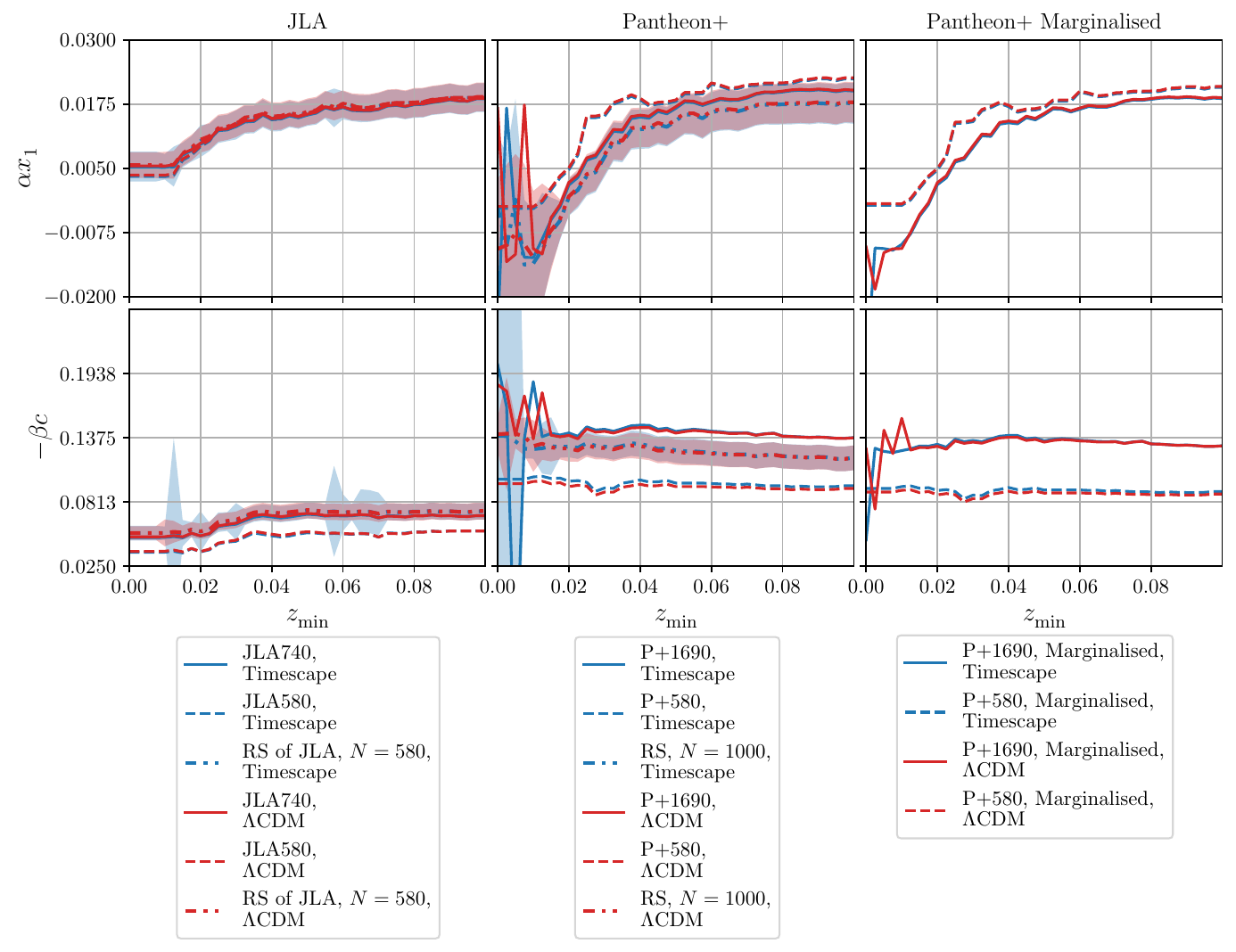}
    \caption{We present additional panels to \cref{fig:random} that depict the contributions of the $\alpha x\Z 1$ and $-\beta c$ terms to the distance modulus $\mu$. The solid lines and shaded regions indicate the median and one standard deviation at a given $\zmin$, respectively, calculated from our set of random subsamples of the JLA (left) and \pplus\ (middle) datasets based on the non-marginalised likelihood (\cref{eq:likelihoodFull}) and the frequentist approach. The results in the right column are obtained from the P+1690 sample and the common subsample by considering the marginalisation (\cref{eq:likelihoodMarg}).}
    \label{fig:lightcurveparams_products}
\end{figure*}

These distance moduli refer to ideal observers and an isotropic distance--redshift relation, while the results from the SALT relation (\cref{eq:tripp}) refer to the actual emitter (em) and observer (obs). Transforming \cref{eq:dLlcdm} and
\cref{dLTS} to the frame using the actually measured
redshift $\zh=(\lambda\ns{obs}-\lambda\ns{em})/\lambda\ns{em}$ gives
\beq
\hat\dL(\zh)=\frac{1 + \zh}{1 + z} \dL(z)=(1+\zh)D(z),
\eeq
to be entered in \cref{eq:mu}. Here, $D(z)=\dL/(1+z)=(1+z)\dA$ is the
(effective) comoving distance for each cosmological model, and
\beq
1+\zh=(1 + z)(1+z^{\rm pec}\ns{obs})(1+z^\phi\ns{obs})
(1+z^{\rm pec}\ns{em})(1+z^\phi\ns{em})\label{redsplit}
\eeq
gives the measured redshift, $\zh$, (heliocentric redshift in our case) in terms of the cosmological redshift, $z$,
the local Doppler redshifts of observer, $z^{\rm pec}\ns{obs}$, and emitter,
$z^{\rm pec}\ns{em}$, and gravitational redshifts at the two locations,
$z^\phi\ns{obs}$ and $z^\phi\ns{em}$. As discussed by \citet{Dam_2017}, we calculate the cosmological luminosity distances in the CMB rest-frame without peculiar velocity corrections to the $z^\phi\ns{em}$ term (\cref{sec:input}) and perform our analysis based on these values.

\begin{figure}
    \centering
	\includegraphics[width=0.99\columnwidth]{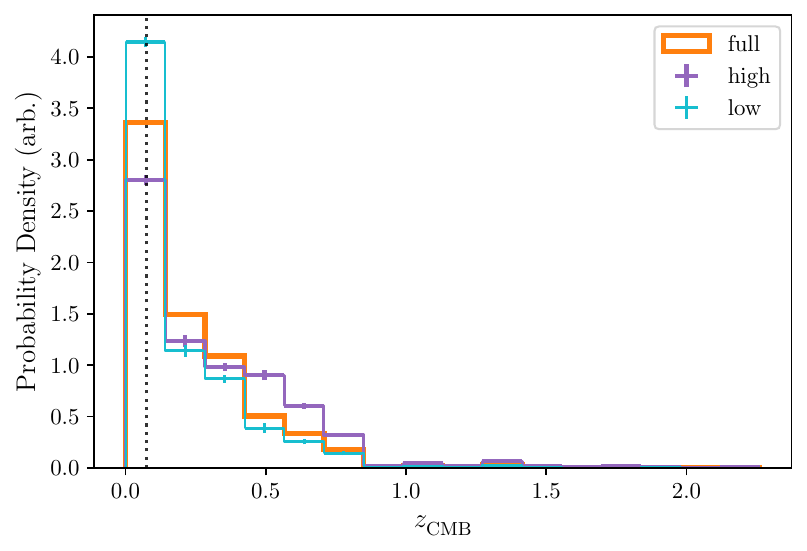}
    \caption{Redshift distribution of the full P+1690 sample and the `high' and `low' subsamples thereof. As described in \cref{sec:commonsubsample}, they are constructed to include more high- and low-redshift \sne, respectively, and are of $N = 580$.}
    \label{fig:TEST}
\end{figure}

The theoretical distance moduli (\cref{muTS,mulcdm}) can be compared to the observational distance moduli given by the Tripp equation (\cref{eq:tripp}). 
However, the parameters $\alpha$ and $x\Z 1$ as well as $\beta$ and $c$ are degenerate, only their products contribute to the distance modulus $\mu$.
In \cref{fig:lightcurveparams_products}, $\alpha x\Z 1$ and $-\beta c$ are plotted for reference purposes. 
As discussed in \cref{sec:shs}, $\beta$ and $c$ 
are subject to Malmquist bias (\cref{fig:ext_z_c}) and are thus not suitable for the definition of SHS$_\beta$. Instead, we use the term $\alpha x\Z 1$ to define SHS$_\alpha$ as the changes in $\alpha$ and $x\Z 1$ balance each other (\cref{fig:ext_z_x1}). 
Furthermore, the light-curve parameters depend on the sample size and the redshift distribution, as discussed in \cref{sec:random}. We analyse this using subsamples with redshift distributions biased towards high and low redshifts. For the resulting redshift distributions see \cref{fig:TEST} and \cref{fig:conv} for the light-curve parameters.

\newpage
\section{Data for the Statistical Analysis}\label{app:statdata}

For the statistical analysis, we calculate a covariance matrix for the \pplus\ dataset as described in \cref{sec:input}. The full sample contains $\mathcal{N} = 1708$ supernova observations (including duplicates), each of which contributes a $3\times3$ block to the statistical part of the covariance pertaining to the light-curve parameters. However, as detailed in \cref{sec:statcov}, certain covariance matrices of individual supernovae have negative eigenvalues and were consequently dropped from our sample. In instances where this occurred in one survey, the corresponding datasets for all surveys featuring the relevant supernova were omitted. \cref{tab:dropped} provides a comprehensive record of the \sne\ eliminated by this procedure.
The resulting sample consists of $\mathcal{N} = 1690$ supernova observations and its full covariance matrix is positive semi-definite.

The light-curve parameters and the covariance matrix are then used for both the frequentist and the Bayesian analysis (\cref{sec:statmethods}). For the Bayesian approach, we adopted the same priors as \citet{Dam_2017} (except with a larger $\beta$ range), which are listed in \cref{tab:priors}. Each parameter prior was assumed to have a uniform distribution, while the standard deviations are uniformly logarithmic, such as was done in the analyses of \citet{March_2011}, \citet{Nielsen_2016}, and \citet{Dam_2017}. For the model parameters, \citet{Dam_2017} constructed the priors based on $2\sigma$ bounds from CMB and BAO observations, as detailed in \citet[Appendix~D]{Dam_2017}. The resulting intervals which were used as priors are given in \cref{tab:priors}.

\onecolumn
\begin{table}
\centering
\caption{The supernovae excluded from our analysis, as detailed in \cref{sec:statcov}, are those that exhibit negative eigenvalues in any of their observations, rendering their light-curve parameter covariances not positive semi-definite. It is worth noting that the values of $z\Ns{CMB}$ reported in this study are based on the boost method outlined in \cref{sec:input}.}
\begin{tabular}{llrrrrrrrcc}
\hline 
Supernova & Survey &  $z\Ns{CMB}$ &  $z\Ns{Helio.}$ &      RA &     DEC &  $m\Ns{B}$ &  $x\Z 1$ &    $c$ &  Negative Eigenvalues &   JLA \\  %[1ex] 
\hline
\hline
100358 & PS1MD &          0.366 &            0.368 & 352.444 &   0.591 & 21.779 & -0.430 & -0.057 &                     1 &       \\
120400 & PS1MD &          0.352 &            0.350 & 149.628 &   2.872 & 21.773 &  0.187 & -0.121 &                     1 &       \\
120444 & PS1MD &          0.299 &            0.300 &  51.682 & -27.936 & 21.315 &  0.277 & -0.186 &                     2 &       \\
470041 & PS1MD &          0.332 &            0.331 & 131.337 &  44.755 & 21.453 &  0.938 & -0.070 &                     1 &       \\
510266 & PS1MD &          0.441 &            0.440 & 214.033 &  51.732 & 22.045 & -0.476 & -0.170 &                     1 &       \\
550041 & PS1MD &          0.258 &            0.260 & 352.540 &  -0.866 & 20.671 &  0.799 & -0.192 &                     1 &       \\ \hline
Carter & CANDELS &          1.543 &            1.540 & 150.061 &   2.192 & 26.138 &  2.349 &  0.133 &                     1 &       \\ \hline
SDSS12927 & SDSS &          0.189 &            0.190 &  41.786 &   0.775 & 20.077 &  0.095 & -0.034 &                     1 &  True \\
SDSS15234 & SDSS &          0.135 &            0.136 &  16.958 &   0.828 & 20.213 &  0.892 &  0.130 &                     1 &  True \\
SDSS7473 & SDSS &          0.217 &            0.218 &   4.326 &  -0.257 & 20.631 & -0.148 & -0.100 &                     1 &  True \\ \hline
SN2001eh & CFA3S &          0.036 &            0.037 &  24.550 &  41.655 & 16.407 &  1.713 & -0.030 &                     0 &  True \\
          & LOSS2 &          0.036 &            0.037 &  24.550 &  41.655 & 16.381 &  1.786 & -0.043 &                     1 &  True \\ \hline
SN2003kf & CFA3S &          0.008 &            0.007 &  91.148 & -12.629 & 13.312 &  0.525 & -0.021 &                     1 &       \\
SN2008bc & CSP &          0.016 &            0.015 & 144.630 & -63.974 & 14.416 &  0.574 & -0.132 &                     1 &       \\ \hline
SN2009d & CFA4p2 &          0.025 &            0.025 &  58.595 & -19.182 & 15.461 &  0.425 & -0.071 &                     0 &       \\
          & CSP &          0.025 &            0.025 &  58.595 & -19.182 & 15.432 &  0.650 & -0.097 &                     0 &       \\          & LOSS1 &          0.025 &            0.025 &  58.595 & -19.182 & 15.461 &  1.045 & -0.088 &                     1 &       \\ \hline
SN2016hhv & FOUND &          0.061 &            0.062 & 353.952 &  20.709 & 17.618 & -0.771 & -0.182 &                     1 &       \\
\hline
\end{tabular}
\label{tab:dropped}
\end{table}

\begin{table}
\centering
\caption{Bayesian and frequentist prior parameters that were used for the analysis. It should be noted that all priors are uniform on the respective interval and, importantly, relatively broad to account for timescapes' broader conditions.}
\begin{tabular}{c | cc | cc } 
    \hline
     & Timescape & & Spatially flat $\Lambda$CDM & \\
     \hline
     & Freq. & Bayes. Priors & Freq. & Bayes. Priors \\ 
    \hline
    \hline
    $f\Ns{v0}$ & 0.763 & [0.500,0.799] ($2 \sigma$ bound) & & \\ [0.8ex]
    $\Omega\Ns{M0}$ & & & 0.376 & [0.143,0.487] ($2 \sigma$ bound) \\ [0.8ex]
    $\alpha$ & 0.131 & [0,1] & 0.134 & [0,1] \\ [0.8ex] 
    $\beta$ & 3.12 & [0,7] & 3.12 & [0,7] \\ [0.8ex] 
    $x\Z 1$ & 0.0981 & [-20,20] & 0.0995 & [-20,20] \\ [0.8ex] 
    $\sigma_{x\Z 1}^2$ & 0.812 & $\log_{10}(\sigma)$ {[-10,4]} & 0.812 & $\log_{10}(\sigma)$ {[-10,4]} \\ [0.8ex] 
    $c$ & -0.0198 & [-20,20] & -0.0201 & [-20,20] \\ [0.8ex] 
    $\sigma\Ns{c}^2$ & 0.00480 & $\log_{10}(\sigma)$ {[-10,4]} & 0.00479 & $\log_{10}(\sigma)$ {[-10,4]} \\ [0.8ex] 
    $M$ & -19.3 & [-20.3,18.3] & -19.3 & [-20.3,18.3] \\ [0.8ex] 
    $\sigma\Ns{M}^2$ & 0.0107 & $\log_{10}(\sigma)$ {[-10,4]} & 0.0107 & $\log_{10}(\sigma)$ {[-10,4]} \\ [0.8ex] 
    \hline
\end{tabular}
\label{tab:priors}
\end{table}

\renewcommand\thefigure{C\arabic{figure}}
\begin{figure*}
    \centering
	\includegraphics[width=\textwidth]{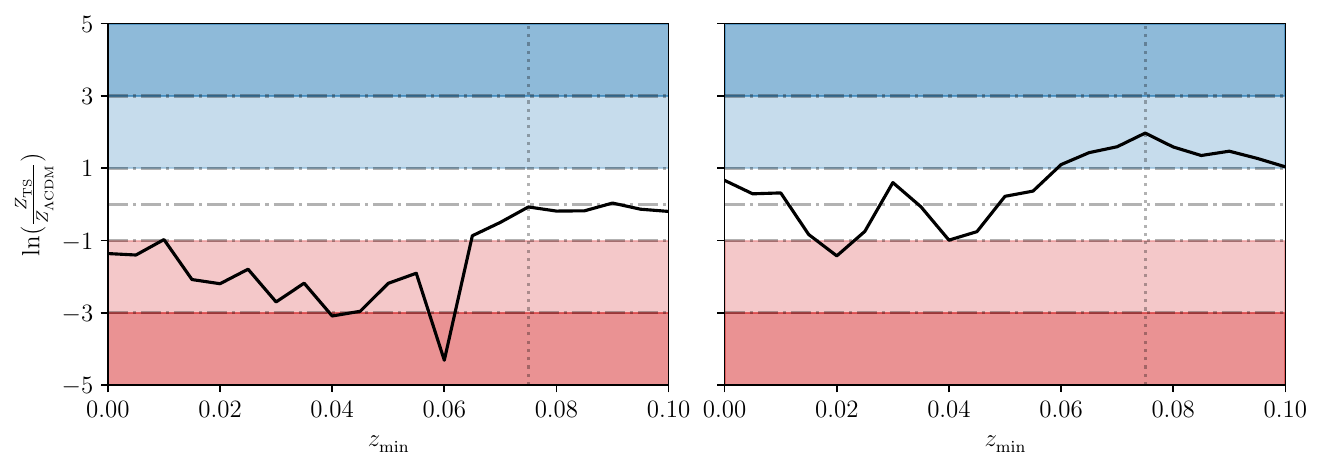}
    \caption{{{\bf(a)} Repeat analysis for the P1690+ panel of \cref{fig:modelcomp} including FLRW model peculiar velocity corrections, which are artificial in the case of timescape, to induce a maximum possible \LCDM\ bias. {\bf(b)} Identical FLRW model peculiar velocity corrections applied to the new Tripp method analysis of \citet{Seifert_2024} result in no significant preference for either model.}}
    \label{fig:flrw}
\end{figure*}

\newpage

\begin{multicols}{2}

\section{FLRW Peculiar Velocity Models}\label{app:vpec}

To further explore the impact of the FLRW peculiar velocity corrections on the relative Bayesian evidence, particularly at low redshifts, we undertook an analysis that included the full FLRW peculiar velocity corrections to \emph{both} the FLRW and timescape distance moduli. This analysis used a modified covariance matrix, as described in \cref{sec:input}, and included the Hubble diagram redshift, $z\Ns{HD}$, adjusted for peculiar velocities determined from the FLRW model. The crucial ingredient of the timescape model for our tests is the uniform quasilocal Hubble expansion condition, which holds down to scales of order $3$--$15\,$Mpc. Thus by artificially including FLRW peculiar velocity corrections which remove a key timescape feature, any Bayesian preference for timescape at low redshift should removed, or reversed in favour of the \lcdm\ model. For high redshift cuts, $\zmin$, the contribution of peculiar velocity corrections is small in any cosmological model with large scale statistical isotropy, including both \LCDM\ and timescape. As a result the FLRW peculiar velocity check should have a negligible effect for large $\zmin$.

The above expectations are borne out by the analysis shown in Fig.~\ref{fig:flrw}(a). Relative to the lower P+1690 panel of Fig.~\ref{fig:modelcomp} we see that for the full sample $\ln B$ decreases markedly with nominal Bayesian evidence now in favour of \LCDM\ at a moderate level, with comparable evidence for cuts $\zmin<0.075$. Beyond SHS$_\alpha$ the FLRW peculiar velocity changes have no statistical effect.

Finally, we also applied the FLRW model peculiar velocity check to the refined Tripp statistical method from \citet{Seifert_2024}, which gives evidences $\ln{Z\Ns{Tripp}} \gg \ln{Z\Ns{Gauss}}$. For cuts in the low-redshift nonlinear regime ($\zmin < 0.025$), Fig.~\ref{fig:flrw}(b) shows no significant evidence favouring either model, despite the intrinsic bias towards FLRW in this methodology. This further challenges any potential misconception that incorporating peculiar velocities into the \lcdm\ model will eliminate the timescape model as a viable cosmological alternative.

Naturally, future development of the timescape cosmology to account for small local peculiar velocities is desirable. It must be stressed that redshifts $z\lesssim0.025$ correspond to the nonlinear regime of structure formation, which must be understood via cosmological simulations of Einstein's equations in the case of timescape, and are beyond the scope of this article. Nonetheless, the analysis here is consistent with the small non-FLRW peculiar velocity corrections being subdominant relative to regional nonkinematic peculiar expansion \citep{Williams_2024}.

\end{multicols}

%%%%%%%%%%%%%%%%%%%%%%%%%%%%%%%%%%%%%%%%%%%%%%%%%%
% \the\columnwidth is 244
% Textwidth is 508
% Don't change these lines
\bsp	% typesetting comment
\label{lastpage}
\end{document}